\documentclass[12pt]{article}
\usepackage{amsmath}
\usepackage{graphicx}
\usepackage{enumerate}
\usepackage{natbib}
\usepackage{url}
\usepackage{amsfonts}
\usepackage{amssymb}
\usepackage{caption}
\usepackage{subcaption}
\usepackage{float}
\usepackage[dvipsnames]{xcolor}
\usepackage{hyperref}
\usepackage{bbm}

\newtheorem{theorem}{Theorem}

\newtheorem{assumption}{Assumption}

\newtheorem{corollary}{Corollary}

\newtheorem{example}{Example}

\newtheorem{proposition}{Proposition}
\newtheorem{remark}{Remark}

\input{tcilatex}
\def\spacingset#1{\renewcommand{\baselinestretch}{#1}\small\normalsize}
\small\normalsize
\spacingset{1}
\graphicspath{{C:/Surfdrive/VG2/GHillier/JBES/}}

\begin{document}

\title{\textbf{Improved Tests for Mediation\thanks{This is a \emph{revised} version of the paper by Grant Hillier, Kees Jan van Garderen, Noud van Giersbergen (2022): Improved tests for mediation, cemmap working paper, No. CWP01/22, Centre for Microdata Methods and Practice (cemmap), London, \url{https://doi.org/10.47004/wp.cem.2022.0122}}}}
\author{Grant Hillier\\ CeMMAP and University of Southampton \\
and \\
Kees Jan van Garderen \\
UvA-Econometrics, University of Amsterdam \\
and \\
Noud van Giersbergen \\
UvA-Econometrics, University of Amsterdam}
\maketitle

\begin{abstract}
Testing for a mediation effect is important in many disciplines, but is made
difficult - even asymptotically - by the influence of nuisance parameters.
Classical tests such as likelihood ratio (LR) and Wald (Sobel) tests have
very poor power properties in parts of the parameter space, and many
attempts have been made to produce improved tests, with limited success. In
this paper we show that augmenting the critical region of the LR test can
produce a test with much improved behavior everywhere. In fact, we first
show that there exists a test of this type that is (asymptotically) exact
for certain test levels $\alpha $, including the common choices $\alpha
=.01,.05,.10.$ The critical region of this exact test has some undesirable
properties. We go on to show that there is a very simple class of augmented
LR critical regions which provides tests that are nearly exact, and avoid
the issues inherent in the exact test. We suggest an optimal and coherent
member of this class, provide the table needed to implement the test and to
report p-values if desired. Simulation confirms validity with non-Gaussian
disturbances, under heteroskedasticity, and in a nonlinear (logit) model. A
short application of the method to an entrepreneurial attitudes study is
included for illustration.
\end{abstract}

\bigskip

\newpage

\spacingset{1.9}%

\section{Introduction}

\label{sec:introduction}

Testing for a mediation effect has important applications in many
disciplines, including psychology, sociology, epidemiology, accounting,
marketing, economics and business.\footnote{%
See, for example, \citet{Baron1986}, \citet{Coletti2005}, %
\citet{Mackenzie1986}, \citet{Alwin1975}, \citet{Freedman1992}, %
\citet{Heckman2015a, Heckman2015b}.} A simple context for the problem -
which we will use to motivate the results to follow - is a model of the type%
\begin{eqnarray}
m &=&\theta _{1}x+u_{1},  \label{eq:mx} \\
y &=&\tau x+\theta _{2}m+u_{2}.  \label{eq:yxm}
\end{eqnarray}%
This model is generally given a causal interpretation with $x$ exerting a
causal influence on $y$ via the mediating variable $m$. The influence of $x$
on $y$ may be both direct (the term $\tau x),$ and/or indirect via the term $%
\theta _{2}m$ if $\theta _{1}$ is nonzero in the first equation. There is no
mediation effect if either $\theta _{1}=0$ such that $x$ does not appear in (%
\ref{eq:mx}), or $\theta _{2}=0$ such that $m$ does not appear in (\ref%
{eq:yxm}). So a test for the \emph{absence} of a mediation effect is a test
of the composite null hypothesis $H_{0}:\theta _{1}\theta _{2}=0$. Controls
can be added to the model to avoid unmeasured confounding effects. The
vectors $y,x,$ and $m$ are then the residuals after regression on these
controls. This has no bearing on what follows if standard assumptions
regarding controls are met: after including controls (covariates) no
un-measured common causes exist for the relations between (i) $x$ and $m,$
(ii) $x$ and $y,$ (iii) $m$ and $y$, and (iv) $m$ and $y$ that are affected
by $x$; see for instance \citet[p.26]{vdweele2015}.

This testing problem is complicated by the fact that, even asymptotically,
there is a nuisance parameter present under the null - either $\theta _{1}$
or $\theta _{2}$ may be nonzero - and this seriously impacts the properties
of most of the tests that have been proposed for the problem; see %
\citet{Mackinnon2002} for a survey. Typically, the extant tests have very
poor power behavior near the origin ($\theta _{1}=\theta _{2}=0).$
Specifically, the null rejection probability (NRP) of the test can be very
much smaller than its nominal size - near zero in fact - and its power and
NRP can be very nearly equal. The bank of standard tests exhibiting this
behavior all reject the null hypothesis when some particular test statistic
is large. However, in a recent paper, \citet{VG2-2021}, VG2 hereafter, have
shown that both NRP and power can be improved considerably - particularly
near the origin - by using a critical region that cannot be defined in this
way, but is simply a subset of a two-dimensional sample space. After
reducing the problem by invariance, they consider a critical region $(CR)$
consisting of the likelihood ratio region ($CR_{LR})$, augmented by an
additional region closer to the origin. This additional region is carefully
constructed using a piecewise-linear spline, and is optimized in terms of
both NRP and power.

In this paper we employ the same idea - augmenting the $CR_{LR}$ by an
additional region - but, in the interests of pragmatism, our focus here will
be on constructing a test that is very simple to apply in empirical
research, yet has\ NRP very nearly constant.

To motivate the proposed test we first show that, for certain test sizes
(including the popular choices $\alpha =.01,.05,$ and $.10)$ a test exists
that is exactly similar asymptotically, i.e. with constant NRP equal to $%
\alpha $, and we show how to construct it. The construction of this exact
test resembles those mentioned by \citet[p.542]{lehmann1952testing} and
later by \citet[p.492]{nomakuchi1987note}; see also \citet{berger1989} for
related constructions. As in those earlier examples, however, the critical
regions have some undesirable characteristics including the lack of \textit{%
coherency}. A test procedure is said to be \textit{coherent} if, when the
test rejects the null at level $\alpha $, it also rejects at all levels
greater than $\alpha $. The likelihood ratio test has this property, see %
\citet{VG2-2022}, but we show below that the exact test does not.

We therefore propose an alternative, easily constructed, test that is close
to being exact, and which avoids some of the undesirable aspects of the
exact test. We call this the "simply-augmented LR test" as it adds a region
with linear boundary defined by two points to the $CR_{LR}$. Specifically,
using the two common $t$-statistics for $\theta _{1}$ and $\theta _{2},$ if $%
v_{1}=\min \{t_{1}^{2},t_{2}^{2}\}$ and $v_{2}=\max \{t_{1}^{2},t_{2}^{2}\},$
then the $.05$ level test is simply:\newline
\hspace*{12mm} \emph{reject} $H_{0}:\theta _{1}\theta _{2}=0$ if $v_{1}>$ $%
3.841$ (the LR test) or $v_{1}/v_{2}>0.8744$ (augmentation), \ \newline
i.e. reject if a traditional LR\ test rejects and otherwise check the
augmented region. Only one number is required - in addition to the usual $%
\chi _{1}^{2}$-critical value - in order to implement the test. We provide
these critical values for every percentile level in Table \ref{tab:alfa_b_z}
at the end of the paper. We prove theoretically that this test cannot be
exactly size-correct. It can be modified to ensure that its NRP $\leq \alpha 
$ for all parameter values under the null, but the resulting
(truncated-augmented) test lacks coherency. Coherency is crucial for the
interpretability and derivation of p-values in empirical research. Table \ref%
{tab:alfa_b_z} also shows the straightforward calculation of p-values based
on the coherent simply-augmented LR test. The new test is far superior to
the LR test in terms of NRP, and, trivially (because their critical regions
are larger), also has greater power.

For any test with critical region $w$, and where the distribution of the
statistics involved depends on a vector of parameters $\psi ,$ we denote the
power of the test by $P_{w}(\psi ),$ and the NRP when the null distribution
depends on the parameter $\psi _{0}$ by $P_{w}(\psi _{0}).$ The \textit{size}
of the test is as usual defined to be $\sup_{\psi _{0}}P_{w}(\psi _{0}).$ We
emphasize that the issue we are concerned with here is not that of finding
tests of the correct size - the LR and Sobel's Wald test both have this
property - but that the usual tests can have NRP and power that are near
zero in relevant parts of the parameter space where mediation effect is
small or imprecisely estimated.

Section \ref{sec:model_testing_invariance} motivates restricting our search
for an improved testing procedure to the squared $t$-statistics. We
formulate the asymptotic problem under minimal assumptions that allow for\
nonnormality and heteroskedasticity, which is important empirically. Section %
\ref{sec:LRtest} analyzes the LR test and shows the poor NRP and power
properties, motivating the search for an improved test. Section \ref%
{sec:exact_test} proves the existence of an exact, but incoherent test.
Section \ref{sec:simpler_augmented_LR_tests} considers the augmented tests
and derives their properties. Section \ref%
{sec:power_coherence_p_values_example} shows the coherency of the
simply-augmented test and how this leads to p-values. Section \ref%
{sec:Simulations} shows the relevance of the (asymptotic-based) procedures
in finite samples encountered in practice by simulation and an empirical
application before concluding in Section \ref{sec:conclusion}.

\section{The testing problem}

\label{sec:model_testing_invariance}

The testing problem for $H_{0}:\theta _{1}\theta _{2}=0$ respects several
symmetries, such as the signs of the coefficients. Furthermore, the truth or
falsity of the null is irrespective of the error variances, or the value of $%
\tau $. We are looking for testing procedures that are invariant to these
transformations. They can be based solely on the ordinary $t$-ratios for $%
\theta _{1}$ and $\theta _{2}$ since they are shown to be the maximal
invariants under the relevant group of transformations in Theorem A.1 of the
Appendix under Gaussianity. The LR test and Sobel's Wald test are in fact
basic functions of only these two statistics. The Gaussian assumption is
unnecessarily restrictive however. The two $t$-statistics converge to
normality under the much weaker conditions based on \citet{White1980}, that
we introduce next. Heteroskedasticity in particular is ubiquitous in
empirical research, and can be addressed using robust $t$-statistics.
Proposition \ref{prop:asymp_distr_T} shows the convergence of the two
relevant robust $t$-statistics defined in Equation (\ref{eq:robust_tstats})
below to a normal distribution under the following assumptions.

\begin{assumption}
\label{assump:1} \ \newline
\vspace{-3.0em}

\begin{description}
\item[(i)] $\mathbf{Y}_{i}=\mathbf{X}_{i}^{\prime }\mathbf{\mathbf{\beta }}+%
\mathbf{E}_{i},\qquad i=1,...,n,$ \newline
where $\mathbf{Y}_{i}=\left( m_{i},y_{i}\right) ^{\prime },$ $\mathbf{E}%
_{i}=\left( u_{1,i},u_{2,i}\right) ^{\prime }$ and $\mathbf{\beta }=\left(
\theta _{1},\tau ,\theta _{2}\right) ^{\prime }$ and $\mathbf{X=}\left( 
\mathbf{X}_{1}\mathbf{,\cdots ,X}_{n}\right) ^{\prime }$ with $\mathbf{X}%
_{i}^{\prime }=\left( 
\begin{array}{lll}
x_{i} & 0 & 0 \\ 
0 & x_{i} & m_{i}%
\end{array}%
\right) ;$

\item[(ii)] $\{(m_{i},x_{i},u_{1,i},u_{2,i})\}_{i=1}^{n}$ is an independent
sequence;

\item[(iii)] $E[u_{1,i}|x_{i}]=0$, $E[u_{2,i}|x_{i},m_{i}]=0$;

\item[(iv)] for some constants $\delta ,\Delta >0$ and all $i=1,...,n:$ $%
a\in \left\{ 0,2\right\} ,b,c\geq 0,$ $a+b+c=4$ 
\begin{equation}
E\left[ \left\vert u_{1,i}^{a}u_{2,i}^{b}x_{i}^{c}\right\vert ^{1+\delta }%
\right] <\Delta <\infty ;  \label{eq:momentrestrictions}
\end{equation}

\item[(v)] $\Omega _{n}=\frac{1}{n}\sum_{i=1}^{n}E\left[ \mathbf{X}_{i}%
\mathbf{E}_{i}\mathbf{E}_{i}^{\prime }\mathbf{X}_{i}^{\prime }\right] $ is
positive definite for all $n$ sufficiently large;

\item[(vi)] $Q_{n}=E[\frac{1}{n}\mathbf{X}^{\prime }\mathbf{X}]$ is
uniformly positive definite.
\end{description}
\end{assumption}

The essential conditions are that the model defined by the regressions (\ref%
{eq:mx}) and (\ref{eq:yxm}) are correct and therefore that $u_{1}$
conditional on $x$ and $u_{2}$ conditional on $\left( x,m\right) $ have
expectations zero. By the law of iterated expectations this implies that the
disturbances $u_{1}$ and $u_{2}$ are \emph{uncorrelated}, as is usually
assumed explicitly. Observations are assumed to be independent and
sufficient higher-order moments should exist. Note that one of the variables
in $\mathbf{X}$ is $m$ which satisfies Equation (\ref{eq:mx}). Assumptions
can therefore be imposed on $x$ and $u_{1},$ rather than $m$ itself. Further
note that Assumption \ref{assump:1} (iv) implies that the elements of $%
\Omega _{n}$ are uniformly bounded and together with (v) ensure uniform
boundedness of $\Omega _{n}^{-1}$ (see \citet[p.819]{White1980}). Similarly
for $Q_{n}^{-1}.$

\begin{proposition}
\label{prop:asymp_distr_T} Given Assumption \ref{assump:1}, the asymptotic
joint distribution of the robust t-statistics as $n\rightarrow \infty $
satisfies:%
\begin{equation}
(T-\mu )\overset{d}{\rightarrow }N(0,I_{2}),  \label{eq:asympdistr_t}
\end{equation}%
where%
\begin{equation}
T=\left( 
\begin{array}{c}
\sqrt{n}\left[ \hat{D}_{n}^{-1/2}\right] _{11}\hat{\theta}_{1} \\ 
\sqrt{n}\left[ \hat{D}_{n}^{-1/2}\right] _{33}\hat{\theta}_{2}%
\end{array}%
\right) ,  \label{eq:robust_tstats}
\end{equation}%
and 
\begin{equation*}
\mu =\left( 
\begin{array}{l}
\mu _{1} \\ 
\mu _{2}%
\end{array}%
\right) =\left( 
\begin{array}{l}
\sqrt{n}\left[ D_{n}^{-1/2}\right] _{11}\mathbf{\mathbf{\theta }}_{1} \\ 
\sqrt{n}\left[ D_{n}^{-1/2}\right] _{33}\mathbf{\mathbf{\theta }}_{2}%
\end{array}%
\right)
\end{equation*}%
with 
\begin{equation}
D_{n}=Q_{n}^{-1}\Omega _{n}Q_{n}^{-1}\text{ and }\hat{D}_{n}=(\tfrac{1}{n}%
\mathbf{X}^{\prime }\mathbf{X})^{-1}\left( \tfrac{1}{n}\sum_{i=1}^{n}\mathbf{%
X}_{i}\mathbf{\hat{E}}_{i}\mathbf{\hat{E}}_{i}^{\prime }\mathbf{X}%
_{i}^{\prime }\right) (\tfrac{1}{n}\mathbf{X}^{\prime }\mathbf{X})^{-1}.
\label{eq:as_cov_beta}
\end{equation}
\end{proposition}

The remainder of the paper will be based on asymptotic distribution (\ref%
{eq:asympdistr_t}) of the robust $t$-statistics. The problem then becomes:
we observe independent random variables $T_{1},T_{2},$ with $T_{i}\sim N(\mu
_{i},1),i=1,2,$ and wish to test the hypothesis $H_{0}:\mu _{1}\mu _{2}=0.$
It is clear that this problem is invariant under the group of sign changes
\thinspace $T_{i}\mapsto -T_{i}$, $i=1,2,$ and under this group of
transformations the statistics $f_{i}=T_{i}^{2}$, $i=1,2,$ are maximal
invariants. These are independent noncentral $\chi _{1}^{2}$ variates with
noncentrality parameters $\lambda _{i}=\mu _{i}^{2}$, $i=1,2.$ Therefore the
no-mediation restriction $\theta _{1}\theta _{2}=0$ versus $\theta
_{1}\theta _{2}\neq 0$ is equivalent to testing%
\begin{equation}
H_{0}:\min \{\lambda _{1},\lambda _{2}\}=0\text{ against }H_{1}:\lambda
_{i}>0\text{, with }\lambda _{i}=\mu _{i}^{2}\ \text{for }i=1,2.
\label{eq:hypotheses}
\end{equation}%
This problem is clearly also invariant under the group of permutations of $%
(f_{1},f_{2}),$ and maximal invariants under this action are $(v_{1},v_{2}),$
with $v_{i}=f_{(i)}$ the $i^{th}$ order statistic (so $v_{2}\geq v_{1}\geq
0).$ Thus, we are finally led to focus attention on the pair of order
statistics $(v_{1},v_{2})=(f_{(1)},f_{(2)}),$ which live on the octant $%
V=\{(v_{1},v_{2});0\leq v_{1}\leq v_{2}<\infty \}.$ The reader should bear
in mind, though, that any test i.e. $CR$ formulated in terms of $%
(v_{1},v_{2})$ can equally well be re-expressed in terms of the $t$%
-statistics $(T_{1},T_{2})$; see Figure \ref{fig:TversusV}.

\begin{remark}
Although this asymptotic version of the problem has been derived in the
context of the model (\ref{eq:mx}) - (\ref{eq:yxm}), other models may also
lead to an asymptotic testing problem of this form. For example, the
nonlinear (logit) model used in the simulation section.
\end{remark}

The null hypothesis is composite and involves a nuisance parameter $\lambda
=\max \{\lambda _{1},\lambda _{2}\},$ the (possibly non-vanishing)
noncentrality parameter. It is therefore not obvious how to construct a test
(critical region) whose NRP $P_{w}(\lambda )$ does not depend on $\lambda .$
However, we will show below that for each level $\alpha =$ $(r+2)^{-1}$,
where $r$ is a non-negative integer, there exists an exact similar test.
Trivially, these tests have power functions uniformly above that of the LR
test.

As already remarked, the NRP and power of the LR test, and other standard
tests, can in fact be extremely small - NRP when the nuisance parameter is
small, power when both $\lambda _{1}$, $\lambda _{2}$ are small. The popular
Sobel (Wald) test is uniformly much worse than the LR test in both respects
and we will therefore not discuss it further.\footnote{%
\ The Wald test rejects when $W=\frac{v_{1}v_{2}}{v_{1}+v_{2}}$ is large,
using the same critical value as the LR test, but $LR>W.$ The LR is less
biased and more powerful. The $W$ power function stays closer to zero longer
as $\lambda $ increases. If $\lambda =0$ and $\alpha =.05,$ the Wald NRP is $%
.00009$, while $(.05)^{2}=.0025$ for LR test.} There is clearly an incentive
to seek a test whose NRP is closer to the nominal size for all values of the
nuisance parameter, and has better power. This is the motivation for what
follows, which builds on the LR test that we discuss next.

\section{The Likelihood Ratio test and its properties}

\label{sec:LRtest}

In this section we derive the NRP and power of the likelihood ratio (LR)
test. This requires the joint distribution of the maximal invariants, the
order statistics $(v_{1},v_{2})$ that play a central role in the derivation
and the properties of proposed improved tests. The LR test is derived by
minimizing $(t_{1}-\mu _{1})^{2}+(t_{2}-\mu _{2})^{2}$ subject to the
constraint $\mu _{1}\mu _{2}=0.$ It is straightforward to show that this
results in the following critical region in the space of the order
statistics $(v_{1},v_{2})$: reject $H_{0}:\mu _{1}\mu _{2}=0$ when%
\begin{equation}
LR=\min \{f_{1},f_{2}\}=v_{1}
\end{equation}%
is large. As usual, the LR test embodies all invariance properties of the
testing problem. The critical region for the LR test of nominal size $\alpha 
$ is given by the set $CR_{LR}=\{\chi _{\alpha }^{2}<v_{1}<v_{2},v_{2}>\chi
_{\alpha }^{2}\}$ with $\chi _{\alpha }^{2}$ the $\alpha $ critical value
from the $\chi _{1}^{2}$ distribution. Let $g(v;\lambda )$ and $G(v;\lambda
) $ denote the pdf and CDF of the noncentral $\chi _{1}^{2}\left( \lambda
\right) $ distribution. The joint distribution of $(v_{1},v_{2})$ as used
for all subsequent results in the paper equals:%
\begin{equation}
pdf(v_{1},v_{2}|\lambda _{1},\lambda _{2})=\left[ g(v_{1};\lambda
_{1})g(v_{2};\lambda _{2})+g(v_{2};\lambda _{1})g(v_{1};\lambda _{2})\right]
,\ \text{for}\ 0\leq v_{1}\leq v_{2},  \label{eq:pdf_order_stats}
\end{equation}%
based on the premise that $T\sim N(\mu ,I_{2})$; see Proposition A.1 in
Appendix A and its specialization to the null case. Given this distribution,
the following proposition provides a very direct description of the
properties of the LR test.

\begin{proposition}
\label{prop:NRP_LR} Given (\ref{eq:pdf_order_stats}), the NRP $%
P_{CR_{LR}}(\lambda )$ of the LR test of $H_{0}:\min \{\lambda _{1},\lambda
_{2}\}=0$ as a function of $\lambda =\max \{\lambda _{1},\lambda _{2}\}$ is
given by 
\begin{equation}
P_{CR_{LR}}(\lambda )=\Pr [v_{1}>\chi _{\alpha }^{2}|\lambda ]=\alpha
\lbrack 1-G(\chi _{\alpha }^{2};\lambda )],
\end{equation}%
with $\chi _{\alpha }^{2}$ defined by $G(\chi _{\alpha }^{2})=1-\alpha .$\
The power function of this LR test is given by 
\begin{equation}
P_{CR_{LR}}(\lambda _{1},\lambda _{2})=[1-G(\chi _{\alpha }^{2};\lambda
_{1})][1-G(\chi _{\alpha }^{2};\lambda _{2})]=\Pr [f_{1}>\chi _{\alpha
}^{2}]\Pr [f_{2}>\chi _{\alpha }^{2}],
\end{equation}%
with $f_{i}\sim \chi _{1}^{2}(\lambda _{i}),i=1,2.$ This power, $%
P_{CR_{LR}}(\lambda _{1},\lambda _{2})$, will always be greater than its 
\textit{NRP}, $P_{CR_{LR}}(\lambda ),$ since, for given $(\lambda
_{1},\lambda _{2}):$ \ 
\begin{equation*}
\Pr [f_{1}>\chi _{\alpha }^{2}]\Pr [f_{2}>\chi _{\alpha }^{2}]>\min \left[
\alpha \Pr [f_{1}>\chi _{\alpha }^{2}],\alpha \Pr [f_{2}>\chi _{\alpha }^{2}]%
\right] =P_{CR_{LR}}(\lambda ).
\end{equation*}
\end{proposition}

Since, for any fixed $z>0,$ $1-G(z;\lambda )$ is an increasing function of $%
\lambda ,$ tending to one as $\lambda \rightarrow \infty $, we arrive at the
following corollary:

\begin{corollary}
Given (\ref{eq:pdf_order_stats}), the LR test of nominal size $\alpha ,$ and
all $\lambda \geq 0,$ 
\begin{equation}
\alpha ^{2}\leq P_{CR_{LR}}(\lambda )\leq \alpha .
\end{equation}
\end{corollary}

Thus, the LR test has a size of $\alpha $. However, for small $\lambda ,$
the NRP of the LR test can be as small as $\alpha ^{2}$, and it only
approaches the nominal size $\alpha $ as $\lambda \rightarrow \infty $.

The tests we consider below are constructed by augmenting the $LR$ critical
region, and it is clear from the expression for $P_{CR_{LR}}(\lambda )$
above that the region added should have null content either exactly equal to 
$\alpha G(\chi_{\alpha }^{2};\lambda )$ for all $\lambda ,$ rendering the
test exact, or have this property approximately. Both exact and approximate
augmented LR tests will be constructed below.

\section{An exact test}

\label{sec:exact_test}

The poor NRP and power properties of the classical tests motivate the search
for more satisfactory tests. Specifically, we would hope to be able to
construct tests whose NRP is $\alpha $, or nearly so, for all $\lambda ,$
and whose power improves on that of the LR test, in particular. In this
section we shall show that an exact test does indeed exist for certain
choices of $\alpha ,$ and is easily constructed. We confine attention to
tests whose critical regions properly contain that of the LR test. That is,
if $H_{0}$ is rejected by the LR test it must also be rejected by the new
test (but not vice versa).

\begin{figure}[t]
\begin{center}
\includegraphics[width=6in]{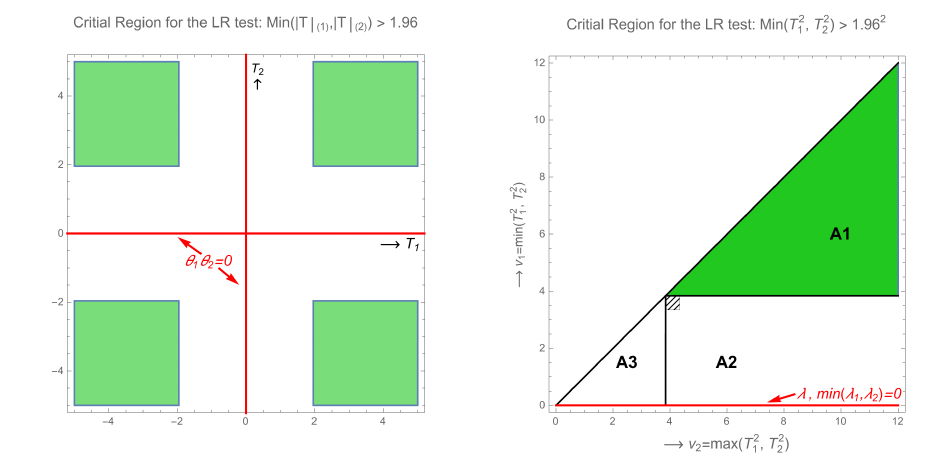}
\end{center}
\par
\vspace*{-6mm}
\caption{The LR rejection regions for $\protect\alpha =.05$ represented by
green colored areas in the sample space for $(T_{1},T_{2})^{\prime }$ and $%
(v_{1},v_{2})^{\prime }$: $\min (|T|_{(1)},|T|_{(2)})>1.96$ in the left
panel and $v_{1}=\min (T_{1}^{2},T_{2}^{2})>1.96^{2}$ in the right panel.
The small rectangle, shaded with lines and contained within $A_2$,
represents the set $R$ from Theorem \protect\ref{Th:no_block_can_b_added}.
Parameter values under the null of no mediation are depicted as red lines.}
\label{fig:TversusV}
\end{figure}

It is convenient for the derivation of the exact test to introduce a
partition of the sample space, the octant $V,$ into three disjoint regions
determined by a scalar $z>0$:$\newline
$ \hspace{6mm} $\ A_{1}=\{v_{2}>z,z<v_{1}<v_{2}\},\
A_{2}=\{v_{2}>z,0<v_{1}<z\},\ A_{3}=\{v_{2}<z,0<v_{1}<v_{2}\}.\newline
$ 
The first of these, $A_{1},$ is the level-$\alpha $ $CR_{LR}$ when $z=\chi
_{\alpha }^{2}$ with acceptance region $AR_{LR}=A_{2}\cup A_{3}$; see Figure %
\ref{fig:TversusV} for a graphical comparison between the CR based on the $t$%
-statistics $t_{1}$ and $t_{2}$ and the CR derived from the order statistics
of the squared $t$-statistics $v_{1}$ and $v_{2}$. In what follows the
regions $A_{1}$, $A_{2}$, $A_{3}$ will be assumed to be defined by $z=\chi
_{\alpha }^{2}$ with probabilities:%
\begin{eqnarray}
P_{A_{1}}(\lambda ) &=&\alpha \lbrack 1-G(\chi _{\alpha }^{2};\lambda )], \\
P_{A_{2}}(\lambda ) &=&1-\alpha -(1-2\alpha )G(\chi _{\alpha }^{2};\lambda ),
\\
P_{A_{3}}(\lambda ) &=&(1-\alpha )G(\chi _{\alpha }^{2};\lambda ), \\
P_{A_{2}\cup A_{3}}(\lambda ) &=&1-\alpha +\alpha G(\chi _{\alpha
}^{2};\lambda ).
\end{eqnarray}

The next proposition determines the NRP of the $CR_{LR},$ i.e. $A_{1}$,
augmented by $A_{3}.$

\begin{proposition}
\label{prop:PA1A3} \ Under the null in (\ref{eq:hypotheses}) and given (\ref%
{eq:pdf_order_stats})\newline
(i)$\ \ \ P_{A_{1}\cup A_{3}}(\lambda )=\alpha +(1-2\alpha )G(z_{\alpha
};\lambda ),$ which varies with $\lambda $ unless $\alpha =1/2.$ \newline
(ii)\ \ For $\alpha =1/2,$ the region $A_{1}\cup A_{3}$ has size $\alpha $
for all $\lambda .$\newline
(iii) For $1/2<\alpha <1,$ $P_{A_{1}\cup A_{3}}(\lambda )<\alpha $ for all $%
\lambda <\infty .$
\end{proposition}

\begin{remark}
Part (ii) shows that there does exist an exact test of size $\alpha =.5,$
namely, the test with CR $A_{1}\cup A_{3}$. We will generalize this property
shortly, and show that exact tests of size $\alpha =(r+2)^{-1}$ exist for
all integers $r\geq 0.$ The case just mentioned is the case $r=0.$
\end{remark}

To illustrate the general result, consider first choosing a single value $%
z_{1}<\chi _{\alpha }^{2}$, with $\alpha $ to be determined also, and using
this to define two disjoint triangular subsets of $A_{3}$:  $%
A_{30}=\{0<v_{1}<v_{2},0<v_{2}<z_{1}\}$ and $A_{31}=%
\{z_{1}<v_{1}<v_{2},z_{1}<v_{2}\,<\chi _{\alpha }^{2}\}$. These have
combined null probability content%
\begin{equation*}
(1-\alpha -G(z_{1}))G(\chi _{\alpha }^{2};\lambda )-(1-\alpha
-2G(z_{1}))G(z_{1};\lambda ),
\end{equation*}%
which differs from the target value $\alpha G(\chi _{\alpha }^{2};\lambda )$
by 
\begin{equation}
(1-2\alpha -G(z_{1}))G(\chi _{\alpha }^{2};\lambda )-(1-\alpha
-2G(z_{1}))G(z_{1};\lambda ).
\end{equation}%
Since $\alpha =1-G(\chi _{\alpha }^{2}),$ we can choose the pair $%
(z_{1},\chi _{\alpha }^{2})$ so that the coefficients of the two noncentral
distribution functions both vanish, yielding a test of size $1-G(\chi
_{\alpha }^{2})$ for \emph{all} $\lambda .$ This requirement produces two
linear equations, $2G(\chi _{\alpha }^{2})-G(z_{1})=1,$ and $G(\chi _{\alpha
}^{2})-2G(z_{1})=0,$ with unique solution $G(z_{1})=1/3,G(\chi _{\alpha
}^{2})=2/3,$ so that $\alpha =1-G(\chi _{\alpha }^{2})=1/3.$ This is the
case $r=1,$ $\alpha =1/3.$

Generalizing this construction, one may prove, as we do in Appendix A:

\begin{theorem}
\label{Th:exacttest}Under the null (\ref{eq:hypotheses}) and given density (%
\ref{eq:pdf_order_stats}) there exist, for each integer $r\geq 0$, unique
numbers $z_{1}<z_{2}<...<z_{r}<\chi _{\alpha }^{2}$ such that the critical
region $CR_{LR}\cup w_{r}(z),$ where $w_{r}(z)=A_{3}\backslash A_{r}(z),$
with 
\begin{equation}
A_{r}(z)=A_{r}(z_{1},...,z_{r})=\bigcup\limits_{i=1}^{r}%
\{0<v_{1}<z_{i},z_{i}<v_{2}<z_{i+1}\},\qquad 
\end{equation}%
where $z_{r+1}=\chi _{\alpha }^{2},$ has null rejection probability $\alpha
=(r+2)^{-1}$ for all $\lambda \geq 0.$ These numbers $z_{i}$ are the
solutions to the identities 
\begin{equation}
G(z_{i})=\frac{i}{r+2},i=1,...,r+1.
\end{equation}%
In particular, $z_{r+1}=\chi _{\alpha }^{2},$ so that $\alpha =(r+2)^{-1}.$
\end{theorem}

When the $z_{i}$ are chosen in this optimal fashion we denote the augmenting
region simply by $w_{r}$.

\begin{figure}[tbph]
\begin{center}
\includegraphics[width=3.5in]{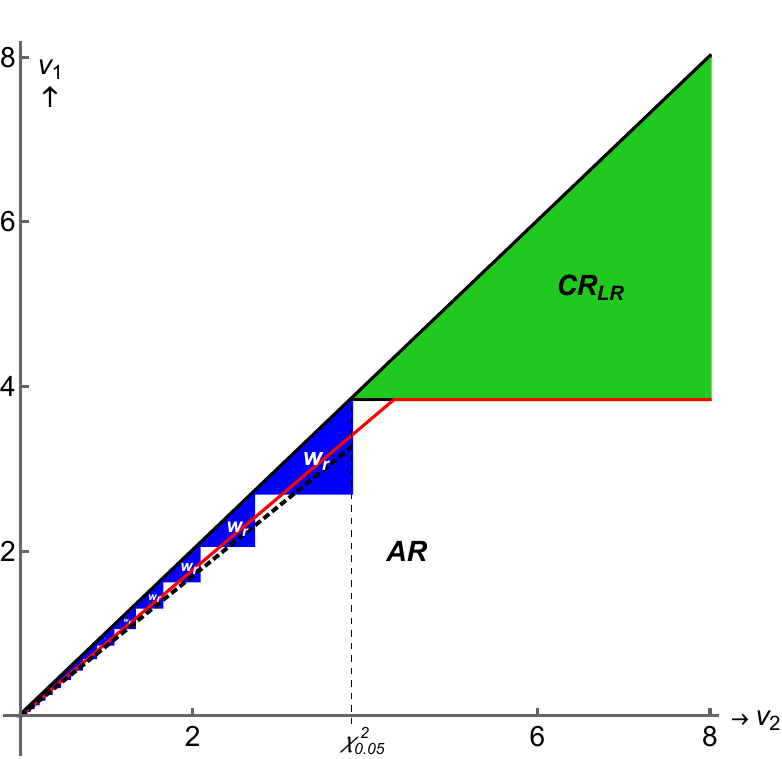}
\end{center}
\par
\vspace*{-6mm}
\caption{$CR_{LR}$ (green) and augmenting region $w_{r}$ (blue) for the
exact test; $\protect\alpha =.05$ (r=18). \newline
Red solid- and black dashed lines are boundaries of augmenting regions, see
Section \protect\ref{sec:simpler_augmented_LR_tests}.}
\label{fig:ExactSimilartest}
\end{figure}

\begin{example}
In the case $\alpha =.05,$ $r=18,$ the critical region is defined by $%
z_{.05}=3.841,$ together with the following 18 values $z_{i}$:\newline

\spacingset{1.0}%

\begin{tabular}{|l|l|l|l|l|l|l|l|l|l|}
\hline
$i\rightarrow $ & $1$ & $2$ & $3$ & $4$ & $5$ & $6$ & $7$ & $8$ & $9$ \\ 
\hline
$z_{i}$ & $.004$ & $.016$ & $.036$ & $.064$ & $.101$ & $.148$ & $.206$ & $%
.275$ & $.357$ \\ \hline
$i\rightarrow $ & $10$ & $11$ & $12$ & $13$ & $14$ & $15$ & $16$ & $17$ & $%
18 $ \\ \hline
$z_{i}$ & $.455$ & $.571$ & $.708$ & $.873$ & $1.074$ & $1.323$ & $1.6424$ & 
$2.0722$ & $2.7055$ \\ \hline
\end{tabular}
\newline

\spacingset{1.9}%

\noindent The augmenting critical region $w_{r}$ is shown in blue in Figure %
\ref{fig:ExactSimilartest} for the case $\alpha =.05$ $(r=18);$ $CR_{LR}$ is
the green region. The solid red line and dashed black lines will be
explained shortly.
\end{example}

\begin{remark}
This construction of exact regions (tests) obviously only works for $\alpha
\leq \frac{1}{2}.$ Using a different constructive proof, VG2 show that,
within a class of tests with weakly increasing cadlag boundary, a similar
test exists if and only if $1/\alpha \in \mathbb{N}$ including $\alpha =1$
and trivially $\alpha =0$. Their test with $\alpha =1/(r+2)$ is essentially
the same as the exact test here.
\end{remark}

\subsection{Power gain}

The power function of the exact test described above is obviously higher
than that of the LR test for all $(\lambda _{1},\lambda _{2}),$ whatever the
value of $r.$ It is easy to check that the probability content of the region 
$A_{3}$ under the alternative is $G(\chi _{\alpha }^{2};\lambda _{1})G(\chi
_{\alpha }^{2};\lambda _{2}).$ The power added by the augmenting region is
therefore given by%
\begin{eqnarray}
P_{w_{r}}(\lambda _{1},\lambda _{2}) &=&G(\chi _{\alpha }^{2};\lambda
_{1})G(\chi _{\alpha }^{2};\lambda _{2})  \notag \\
&&-\sum_{i=1}^{r}\int_{0<v_{1}<z_{i}}\int_{z_{i}<v_{2}<z_{i+1}}\left[
g(v_{1};\lambda _{1})g(v_{2};\lambda _{2})+g(v_{2};\lambda
_{1})g(v_{1};\lambda _{2})\right] dv_{1}dv_{2}  \notag \\
&=&G(\chi _{\alpha }^{2};\lambda _{1})G(\chi _{\alpha }^{2};\lambda
_{2})-\sum_{i=1}^{r}G(z_{i};\lambda _{1})[G(z_{i+1};\lambda
_{2})-G(z_{i};\lambda _{2})]  \notag \\
&&-\sum_{i=1}^{r}G(z_{i};\lambda _{2})[G(z_{i+1};\lambda
_{1})-G(z_{i};\lambda _{1})].
\end{eqnarray}%
This is naturally symmetric in $(\lambda _{1},\lambda _{2}),$ and vanishes
in the limit as either noncentrality parameter goes to infinity. That is,
there is no power gain over the LR test in the limit, but there certainly is
for $(\lambda _{1},\lambda _{2})$ near the origin. The NRP gain at the
origin is obviously $\alpha (1-\alpha )=(r+1)\alpha ^{2}.$ The power
function behaves similarly for points $(\lambda _{1},\lambda _{2})$ close to
the origin with a power gain of around $.0475$ over the LR test when $\alpha
=.05$.

\subsection{Pros and Cons}

For a restricted, but relevant, range of nominal sizes ($\alpha $ of the
form $(r+2)^{-1})$ the construction described above provides, for the first
time, a non-randomized exact test of the no-mediation hypothesis. Whilst the
augmenting critical region does contain points close to the origin, which
might be considered counter-intuitive, over 90\% of the area of the
augmenting critical region in the case of $\alpha =.05$ is accounted for by
the four largest triangular regions, and these regions are well away from
the origin. Nevertheless, the critical region of the exact test does have
several undesirable properties. First, the region $CR_{LR}\cup w_{r}=CR_{r},$
say, is not monotone in $(v_{1},v_{2}).$ That is, $(v_{1},v_{2})\in CR_{r}$
does not imply that $(v_{1}^{\prime },v_{2}^{\prime })\in CR_{r}$ when $%
v_{1}^{\prime }\geq v_{1}$ and $v_{2}^{\prime }\geq v_{2}.$ Similarly, the
acceptance region for the test is not convex, which is somewhat
counter-intuitive. Also, unlike the LR test itself, the exact test does not
possess an important coherence property, namely, that rejection at level $%
\alpha $ does not imply rejection at every level higher than $\alpha .$ That
is, as the reader may easily confirm, the critical region for $\alpha
=(r+2)^{-1}$ is not a subset of that for $\alpha =(r+1)^{-1}.$ This also
rules out the use of p-values. The augmented LR tests introduced in the next
section will address some of these deficiencies.

\section{Simpler augmented LR tests}

\label{sec:simpler_augmented_LR_tests}

To motivate the class of tests we consider next, observe that one could
approximate the exact augmenting region (i.e., the blue $w_{r}$ triangles in
Figure \ref{fig:ExactSimilartest}) with the region above a line $%
v_{1}=bv_{2} $, for some suitable choice of $b$. For instance, the
(geometric) area of the augmenting region of the exact test in the case $%
r=18 $ $(\alpha =.05)$ is $1.08403$, which is equal to the area above the
line $v_{1}=(.87187)v_{2}$ (red solid line in Figure \ref%
{fig:ExactSimilartest}) and below $v_{1}=\chi _{.05}^{2}$.\ Unlike the exact
CR, the NRP of the region defined in this way does vary slightly with $%
\lambda .$ VG2 discusses (in present notation) more general augmenting
regions close to the $v_{1}=v_{2}$ line, but bounded below by a
piecewise-linear spline. The exact test just described is of this form, but
a very special case: the alternate knots are constrained to lie on the line $%
v_{1}=v_{2},$ and the linear components are constrained to be alternately
horizontal and vertical.

Although the triangle above approximates the $w_{r}$ region, it also
includes a region with $v_{2}>\chi _{\alpha }^{2}.$ The next theorem shows
that no $CR_{LR}$ augmented with such a region, nor the CR suggested in VG2,
can be size correct.

\begin{theorem}
\label{Th:no_block_can_b_added}Under the null (\ref{eq:hypotheses}) and
given density (\ref{eq:pdf_order_stats}), if $CR_{LR}$ is augmented by a
non-empty rectangle $R=\{z_{1}<v_{1}<\chi_{\alpha }^{2},\chi_{\alpha
}^{2}<v_{2}<z_{2}\},$ i.e. top left in $A_{2},$ then there is a $\lambda
_{0} $ such that \ $\Pr \left[ CR_{LR}\cup R\right] >\alpha $ for all $%
\lambda _{0}<\lambda <\infty . $
\end{theorem}

This might suggest a truncated version of this simple test with augmentation
region%
\begin{equation*}
w_{b}^{3}=\{(bv_{2}<v_{1})\cap (0<v_{2}<\chi _{\alpha }^{2})\}\subset A_{3},
\end{equation*}%
and critical region $\overline{CR}_{b}=CR_{LR}\cup w_{b}^{3}$, with
superscript indicating that $w_{b}^{3}\ $ is contained in $A_{3}$. We show
that there is an optimal $b,\bar{b},$ for this $\overline{LR}(b)$ test which
is size-correct. We find $\bar{b}$ numerically such that NRP$<\alpha $ for $%
0\leq \lambda \leq \lambda _{0}$, using a relevant $\lambda _{0},$ and for $%
\lambda >\lambda _{0}$ show that for this $\bar{b}$, the NRP is smaller than 
$\alpha $ using the following theorem.

\begin{theorem}
\label{Th:Dtr_smaller0_lambda_inf copy(1)} Given density (\ref%
{eq:pdf_order_stats}) under the null in (\ref{eq:hypotheses}), there exist $%
b<1$ such that $\Pr \left[ \overline{CR}_{b}\right] <\alpha <1$ for all $%
\lambda _{0}<\lambda <\infty .$
\end{theorem}

Like the exact test, the $\overline{LR}(\bar{b})$ test is not coherent,
however, as we show in Section \ref{sec:power_coherence_p_values_example}.

\begin{figure}[h]
\begin{center}
\includegraphics[width=4in]{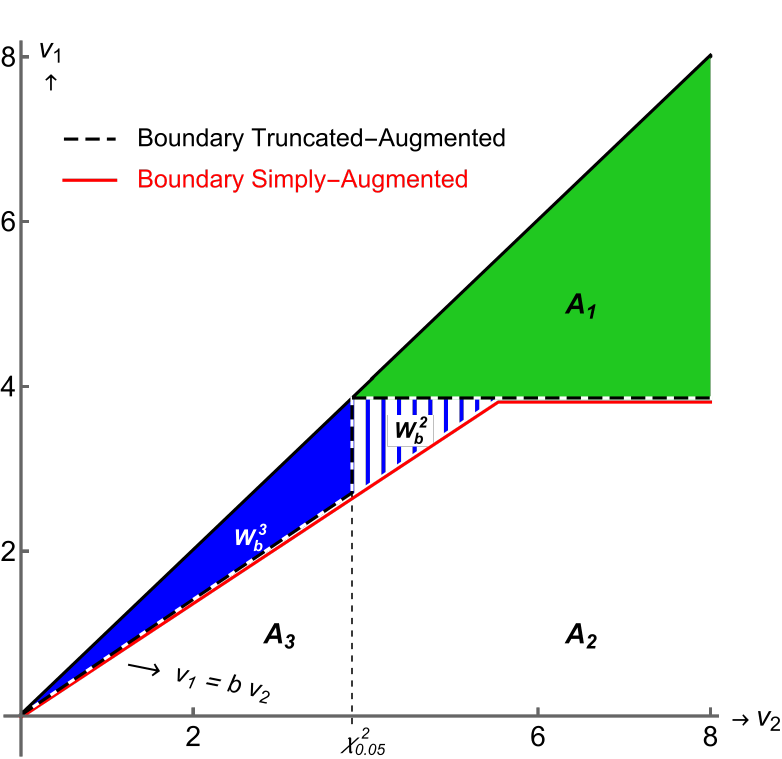}
\end{center}
\par
\vspace*{-3mm}
\caption{Simpler augmented LR tests for $\protect\alpha =0.05$ with $%
w_{b}^{2}\subset A_{2}$ and $w_{b}^{3}\subset A_{3}.$ Truncated-augmented $%
\overline{CR}_{b}=A_{1}\cup w_{b}^{3}$. Simply-augmented $CR_{b}=A_{1}\cup
w_{b}^{2}\cup w_{b}^{3}$. }
\label{fig:CRLRaugment_smpl_trunc}
\end{figure}

\subsection{Simply-augmented LR tests}

\label{sec:simply_augmented_LR_tests}

In view of the earlier comments, we consider the simple augmentation of $%
CR_{LR}$ by the triangular region bounded from below by $v_{1}=bv_{2}$ and
from above by $v_{1}=v_{2}<\chi _{\alpha }^{2}:$%
\begin{equation}
w_{b}=w_{b}^{3}\cup w_{b}^{2}=\{(bv_{2}<v_{1}<v_{2},0<v_{2}<\chi _{\alpha
}^{2})\cup (bv_{2}<v_{1}<v_{2},\chi _{\alpha }^{2}<v_{2}<\chi _{\alpha
}^{2}/b)\}.
\end{equation}%
The test is very simple: reject $H_{0}$ if $v_{1}\geq \chi _{\alpha }^{2}$
or $v_{1}/v_{2}>b$. It can be carried out by doing a LR\ test first. If it
does not reject, then check if the ratio of the two test statistics is
larger than the critical value. We denote the critical regions $CR_{LR}\cup
w_{b}$ as $CR_{b}$ and refer to a test with given value of $b$ as an $LR(b)$
test. Obviously, $LR(1)$ is the LR test. The question to be addressed is how
to choose the appropriate value of $b$. Some properties of this class of
tests that can be used to choose $b$ are given next.

\begin{proposition}
\label{NRP(b)}Under the null (\ref{eq:hypotheses}) and given density (\ref%
{eq:pdf_order_stats}), the NRP of the simply-augmented $LR(b)$ test with
parameter $0<b<1$ is given by%
\begin{eqnarray*}
P_{CR_{b}}(\lambda ) &=&\alpha +A_{\alpha }(b;\lambda )-G(z_{\alpha
};\lambda ), \\
\text{with\ }A_{\alpha }(b;\lambda ) &=&\int_{0<v<\chi _{\alpha
}^{2}}[g(v;\lambda )G(v/b)+g(v)G(v/b;\lambda )]dv.
\end{eqnarray*}%
When $\lambda =0$, i.e. at the origin, 
\begin{equation}
P_{CR_{b}}(0)=2\alpha +2\int_{0<v<\chi _{\alpha }^{2}}g(v)G(v/b)dv-1.
\end{equation}
\end{proposition}

Define the discrepancy function as the difference between the NRP of the
simply-augmented LR test and the nominal value $\alpha $: 
\begin{equation}
D_{\alpha }(b,\lambda )=NRP_{LR(b)}(\lambda )-\alpha =A_{\alpha }(b;\lambda
)-G(\chi _{\alpha }^{2};\lambda ).
\end{equation}%
Since the initial objective was to improve the behavior of the NRP near the
origin, one possibility would be to choose the value of $b$ for which the
test whose NRP is correct at the origin, i.e, the value satisfying $%
D_{\alpha }(b,0)=0$ . This produces a test whose NRP is correct at $\lambda
=0$, and also as $\lambda \rightarrow \infty ,$ but its NRP will be above
the nominal level for intermediate values of $\lambda $.

The following result - partially reiterating Theorem \ref%
{Th:no_block_can_b_added} - says that there is no member of this class of
tests that has NRP equal to $\alpha $ (i.e., $D_{\alpha }(b,\lambda )=0)$
for all $\lambda $:

\begin{proposition}
\label{NonEx}For the simply-augmented $LR(b)$ test, there is no value of $b$
for which $D_{\alpha }(b,\lambda )=0$ for all $\lambda \geq 0.$
\end{proposition}

Now $D_{\alpha }(b,\lambda )$ is obviously continuous in $b$ on the interval 
$(0,1],$ and, for fixed $\lambda ,$ is (strictly) monotonic decreasing in $%
b, $ with $D_{\alpha }(0,\lambda )=(1-2\alpha )G(\chi _{\alpha }^{2};\lambda
)>0 $ when $\alpha <1/2,$ and $D_{\alpha }(1,\lambda )=-\alpha G(\chi
_{\alpha }^{2};\lambda )<0$. This proves the following result:

\begin{proposition}
\label{Uniqueness}For each finite $\lambda $ and $\alpha <1/2$, there is a
unique $b(\lambda )\in \lbrack 0,1]$ for which $D_{\alpha }(b,\lambda )=0$
for the simply-augmented $LR(b)$ test. At this point $D_{\alpha }(b,\lambda
) $ changes sign, from positive to negative.
\end{proposition}

\noindent Table \ref{tab:b_fion_lambda} illustrates the behavior of $%
b(\lambda )$ for a few values of $\lambda $ when $\alpha =.05.\medskip $

\spacingset{1.2}%
\begin{table}[tbh]
\caption{Values of $b(\protect\lambda )$ for various values of $\protect%
\lambda ;\ \protect\alpha =.05.$}
\label{tab:b_fion_lambda}\vspace*{-6mm}
\par
\begin{center}
\begin{tabular}{|l|l|l|l|l|l|l|l|}
\hline
$\lambda \rightarrow $ & 0 & .1 & .5 & 1 & 2 & 5 & 20 \\ \hline
$b(\lambda )$ & .8588 & .8599 & .8634 & .8666 & .8707 & .8743 & .8685 \\ 
\hline
\end{tabular}%
\end{center}
\end{table}
\spacingset{1.9}%
The NRP of $LR(b(\lambda ))$ does not exceed the nominal level $\alpha ,$
and choosing the largest $b(\lambda )$ results in NRP$<\alpha $. But this
should hold for all values, also for\ $\lambda $ values not in Table \ref%
{tab:b_fion_lambda}, even when $\lambda \rightarrow \infty $. The LR test
has this asymptotic property, since $G(\chi _{\alpha }^{2};\lambda
)\rightarrow 0$ as $\lambda \rightarrow \infty ,$ and this property is
shared by all members of the class of $LR(b)$ tests:

\begin{proposition}
\label{Limit}For any fixed $b\in (0,1],$ and any $\alpha ,\lim_{\lambda
\rightarrow \infty }D_{\alpha }(b,\lambda )=0.$
\end{proposition}

So $\lim_{\lambda \rightarrow \infty }P_{CR_{b}}(\lambda )=\alpha $ for any $%
b\in (0,1]$, including $P_{CR_{LR}}(\lambda )$, implying $P_{w_{b}}(\lambda
)\rightarrow 0$ as $\lambda \rightarrow \infty $. Theorem \ref%
{Th:no_block_can_b_added} implies nevertheless that there will exist $%
\lambda $ such that $D_{\alpha }(b,\lambda )>0,$ because for any $b<1$, $%
CR_{b}$ will include an area in $A_{2}$. The probability of this area will
be very small however, if this $\lambda $ is large.

These considerations suggest that a reasonable approach to choosing $b$
would be to select the smallest value for which the maximum discrepancy as $%
\lambda $ varies is bounded (small). This will produce a test with maximum
power, subject to the constraint that overrejection is below the chosen
bound.

\subsection{Optimal b's}

Given that smaller $b$ implies higher power, but a value of $b$ too small
leads to an invalid, over-sized test, we choose the smallest $b$ that is
still (approximately) size correct. This is akin to choosing the largest $%
b(\lambda )$ in Table \ref{tab:b_fion_lambda}. So we choose the smallest $%
b<1 $ such that $D_{\alpha }(b,\lambda )\leq \varepsilon ,$ where $%
\varepsilon $ is a small number that needs to be chosen in the absence of a
theoretical justification. It should be larger than 0 since $\varepsilon =0$
would imply correct size and $b=1,$ which is the LR test. Our choice $%
\varepsilon =10^{-9}$ although theoretically positive, is negligible in
practice and choosing $\varepsilon =10^{-16}$ only changes $b$'s, NRPs, and
power by a small margin.

\begin{figure}[h]
\par
\begin{center}
\includegraphics[width=3.5in]{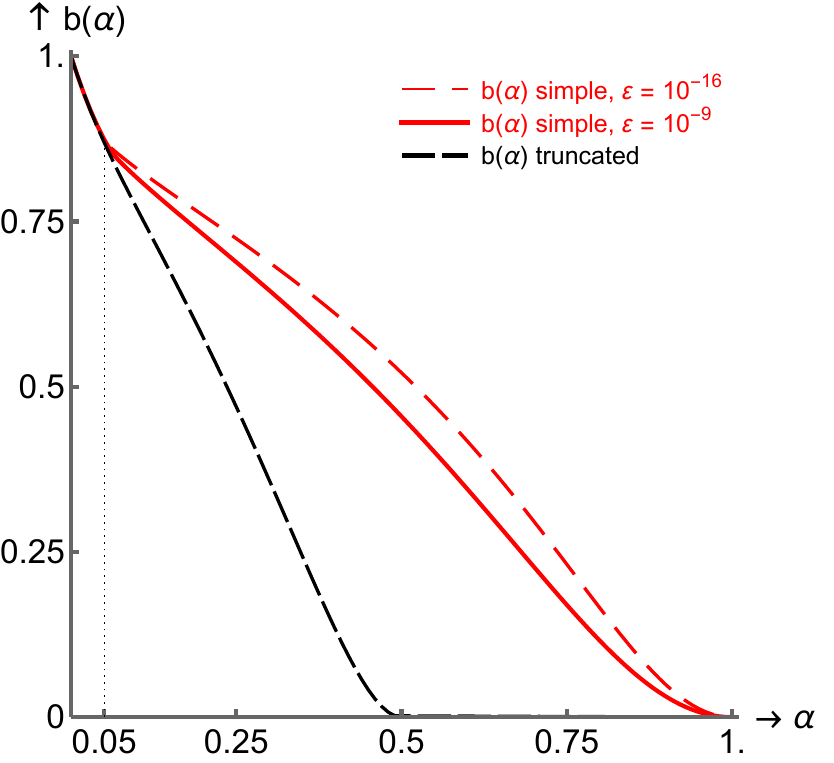}
\end{center}
\par
\vspace*{-6mm}
\caption{$b(\protect\alpha )$ as function of $\protect\alpha $ for the
simpler augmented LR tests: truncated and size correct in dashed black,
simply-augmented in red: solid: over-rejection $<10^{-6}$: dashed: $%
<10^{-16} $.}
\label{fig:optb_fion_of_alfa}
\end{figure}
For each $\alpha $ percentile we determine the optimal (smallest) $b$
numerically such that $D_{\alpha }(b,\lambda )\leq \varepsilon $. Results
are given in Appendix B and shown graphically in Figure \ref%
{fig:optb_fion_of_alfa} for the truncated and the simply-augmented LR tests.
For the simply-augmented LR test numerical values are given in Table \ref%
{tab:alfa_b_z} at the end of the paper.

The optimal $b(\alpha )$ is monotonically decreasing in $\alpha $ for all
three cases. For $\alpha <.05$ there is very little difference between the $%
b $'s. For $\alpha \geq 1/2$ we have $b=0$ for the truncated version that
then has CR $A_{1}\cup A_{3}$ and the NRP $=1/2\geq \alpha $ for all $%
\lambda $.

\subsection{Comparison of RPs under null and alternative}

\label{sec:power}

\begin{figure}[h]
\par
\begin{center}
\includegraphics[width=6in]{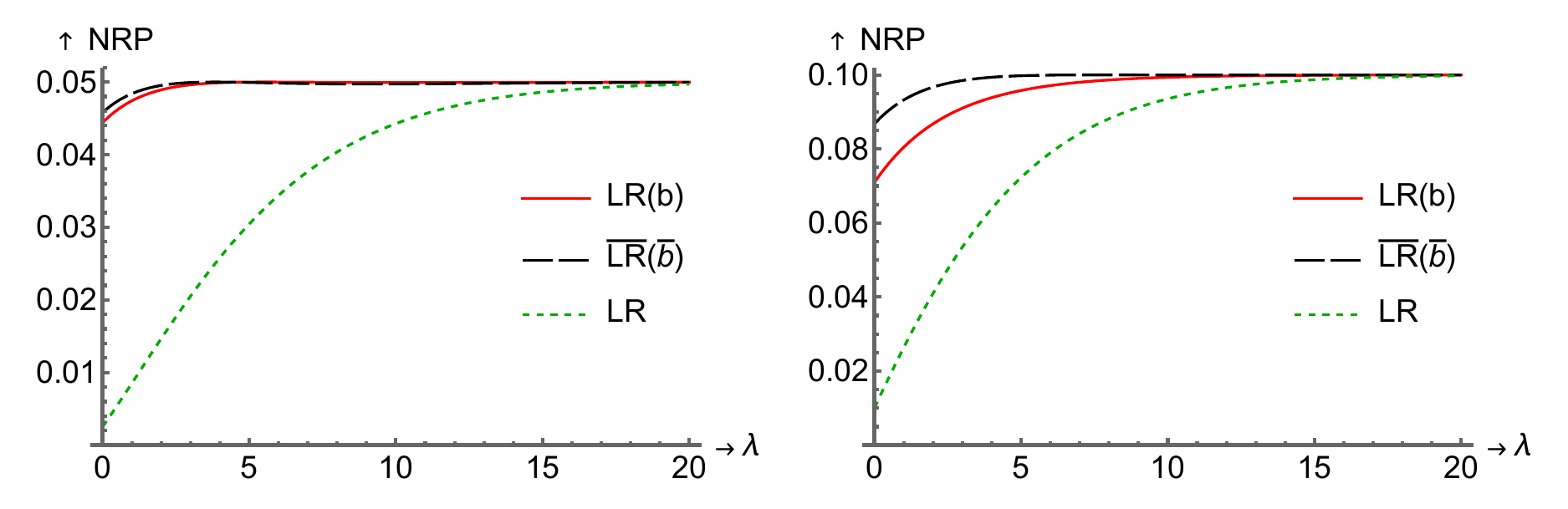}
\end{center}
\par
\vspace{-6mm}
\caption{NRP of LR test and the optimal simply-augmented test; $\protect%
\alpha =.05$ and $.10$}
\label{fig:NRP_LR_LRaug_05_10}
\end{figure}

Figure \ref{fig:NRP_LR_LRaug_05_10} shows the $.05$ and $.10$ level NRPs for
the LR and augmented LR tests. The VG2 test is not displayed, but would show
a horizontal line, deviating from the level $\alpha $ by less than $10^{-9}$
for all $\lambda \geq 0$. 

The LR starts at $\alpha ^{2}$ when $\lambda =0$ and increases to $\alpha $
as $\lambda \rightarrow \infty $. The NRP for $\overline{LR}(\bar{b})$ drops
below that of $LR(b)$ for larger values of $\lambda $ when $w_{b}^{2}$
dominates $w_{\bar{b}}^{3}$ in probability terms. For small values of $%
\lambda $, the NRP of the truncated version is larger, and progressively so
with increasing $\alpha <1/2$. For $\alpha \geq 1/2$, the truncated version
has $\bar{b}=0$ and is the exact similar test when $\alpha =1/2=NRP$ for all 
$\lambda $. \vspace*{6mm}

It is trivially true that the power of the $LR(b)$ and $\overline{LR}(\bar{b}%
)$ tests cannot be less than that of the LR test. The power function of the
augmented LR test can be calculated in exactly the same way as we have done
for the NRP. For the simply-augmented LR test and based on density (\ref%
{eq:pdf_order_stats}) this is: 
\begin{eqnarray}
P(CR_{b}|\lambda _{1},\lambda _{2}) &=&[1-G(\chi _{\alpha }^{2};\lambda
_{1})][1-G(\chi _{\alpha }^{2};\lambda _{2})]-G(\chi _{\alpha }^{2};\lambda
_{1})G(z_{\alpha };\lambda _{2})  \notag \\
&&+\int_{0<v<\chi _{\alpha }^{2}}\left[ g(v;\lambda _{1})G(v/b;\lambda
_{2})+g(v;\lambda _{2})G(v/b;\lambda _{1})\right] dv,  \label{Power}
\end{eqnarray}%
the first term being the power of the LR test. The power function is
obviously symmetric in $(\lambda _{1},\lambda _{2}).$ Again, as either
noncentrality parameter goes to infinity, the power of the augmented test
approaches that of the LR test. The power of the $LR(b)$ and $\overline{LR}(%
\bar{b})$ tests, together with that of the LR test itself (in brackets), are
given in Table \ref{table:PowerLRbLR} for a selection of values of $(\lambda
_{1},\lambda _{2}).$ The table is, of course, symmetric.

\spacingset{1.2}%
\begin{table}[h]
\caption{ Power of various tests: values without superscripts refer to the
simply-augmented test, $\diamond $: truncated-simply-augmented test, $%
\dagger $: LR test, $\ddagger $: VG2 test. All powers are for $\protect%
\alpha =.05$.}
\label{table:PowerLRbLR}\vspace*{-3mm}
\par
\begin{center}
{\footnotesize \ 
\begin{tabular}{|c|c|c|c|c|c|c|}
\hline
$\lambda_{1}$ & $\lambda _{2} \rightarrow $ \ \ .1\ \ \ \ \ \ \  & .5 & 1 & 2
& 5 & 20 \\[0pt] \hline
0.1 & .0454 \hspace{1.0mm} .0038$^\dagger$ &  &  &  &  &  \\ 
& .0467$^{\diamond}$ .0502$^\ddagger$ &  &  &  &  &  \\ \hline
0.5 & .0475 \hspace{1.0mm} .0067$^\dagger$ & .0528 \hspace{1.0mm} .0119$%
^\dagger$ &  &  &  &  \\ 
& .0487$^{\diamond}$ .0511$^\ddagger$ & .0537$^{\diamond}$ .0554$^\ddagger$
&  &  &  &  \\ \hline
1 & .0497 \hspace{1.0mm} .0105$^\dagger$ & .0588 \hspace{1.0mm} .0185$%
^\dagger$ & .0694 \hspace{1.0mm} .0289$^\dagger$ &  &  &  \\ 
& .0507$^{\diamond}$ .0521$^\ddagger$ & .0595$^{\diamond}$ .0605$^\ddagger$
& .0697$^{\diamond}$ .0705$^\ddagger$ &  &  &  \\ \hline
2 & .0530 \hspace{1.0mm} .018$^\dagger$ & .0689 \hspace{1.0mm} .0319$%
^\dagger $ & .0882 \hspace{1.0mm} .0498$^\dagger$ & .124 \hspace{1.0mm} .0858%
$^\dagger $ &  &  \\ 
& .0535$^{\diamond}$ .054$^\ddagger$ & .0692$^{\diamond}$ .0697$^\ddagger $
& .0881$^{\diamond}$ .0888$^\ddagger$ & .1233$^{\diamond}$ .1245$^\ddagger$
&  &  \\ \hline
5 & .0578 \hspace{1.0mm} .0375$^\dagger$ & .0893 \hspace{1.0mm} .0663$%
^\dagger$ & .1287 \hspace{1.0mm} .1035$^\dagger$ & .2052 \hspace{1.0mm} .1784%
$^\dagger$ & .3916 .3706$^\dagger$ &  \\ 
& .0577$^{\diamond}$ .0578$^\ddagger$ & .0888$^{\diamond}$ .0895$^\ddagger$
& .1278$^{\diamond}$ .1291$^\ddagger$ & .2038$^{\diamond}$ .2059$^\ddagger$
& .3898$^{\diamond}$ .3927$^\ddagger$ &  \\ \hline
20 & .0615 \hspace{1.0mm} .0612$^\dagger$ & .1087 \hspace{1.0mm} .1083$%
^\dagger$ & .1695 \hspace{1.0mm} .1691$^\dagger$ & .2918 \hspace{1.0mm} .2912%
$^\dagger$ & .6056 \hspace{1.0mm} .6051$^\dagger$ & .9881 \hspace{ 1.0mm}
.988$^\dagger$ \\ 
& .0614$^{\diamond}$ .0615$^\ddagger$ & .1086$^{\diamond}$ .1087$^\ddagger$
& .1695$^{\diamond}$ .1696$^\ddagger$ & .2917$^{\diamond}$ .2918$^\ddagger$
& .6055$^{\diamond}$ .6057$^\ddagger$ & .9880$^{\diamond} $ .9881$^\ddagger$
\\ \hline
\end{tabular}
}
\end{center}
\end{table}
\spacingset{1.9}%
It is clear that, for $(\lambda _{1},\lambda _{2})$ near the origin, the LR
test has poor power, and that the simply-augmented LR test of size $.05$
improves considerably upon it. The truncated version even more so since it
has smaller $b$, so more area of $A_{3}$ near the origin is added. In
Appendix C we display the power difference $P_{CR_{b}}(\lambda _{1},\lambda
_{2})-P_{CR_{LR}}(\lambda _{1},\lambda _{2})$ for values of the $\lambda
_{i}\in \lbrack 0,10].$ It is evident that the power difference is quite
small for large $\lambda _{i},$ but substantial for $(\lambda _{1},\lambda
_{2})$ near the origin.

We also show $P_{CR_{b}}(\lambda _{1},\lambda _{2})-P_{\overline{CR}_{\bar{b}%
}}(\lambda _{1},\lambda _{2})$ in Appendix B, which is much closer to zero
and implies that the truncation restriction has little effect on the power.
But the price being paid is lack of coherency which we will discuss next.

\section{Coherence and p-values}

\label{sec:power_coherence_p_values_example}

\subsection{Coherence}

\begin{figure}[h]
\centering
\begin{tabular}{cc}
\begin{subfigure}{0.5\textwidth} \centering
\includegraphics[width=0.975\textwidth]{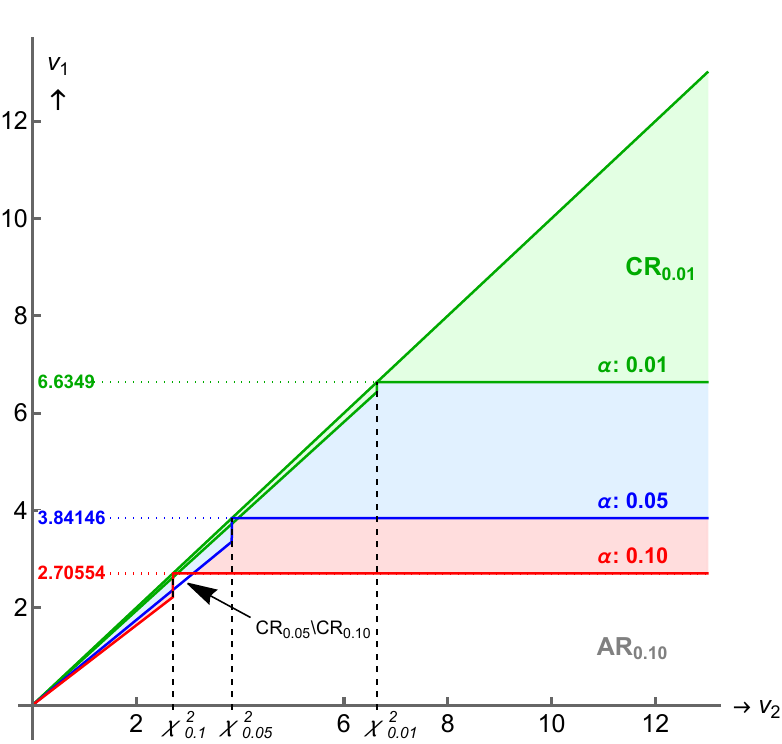}
\caption{CR Truncated simply-augmented} \label{fig:CRtrunc} \end{subfigure}
& \begin{subfigure}{0.5\textwidth} \centering
\includegraphics[width=0.975\textwidth]{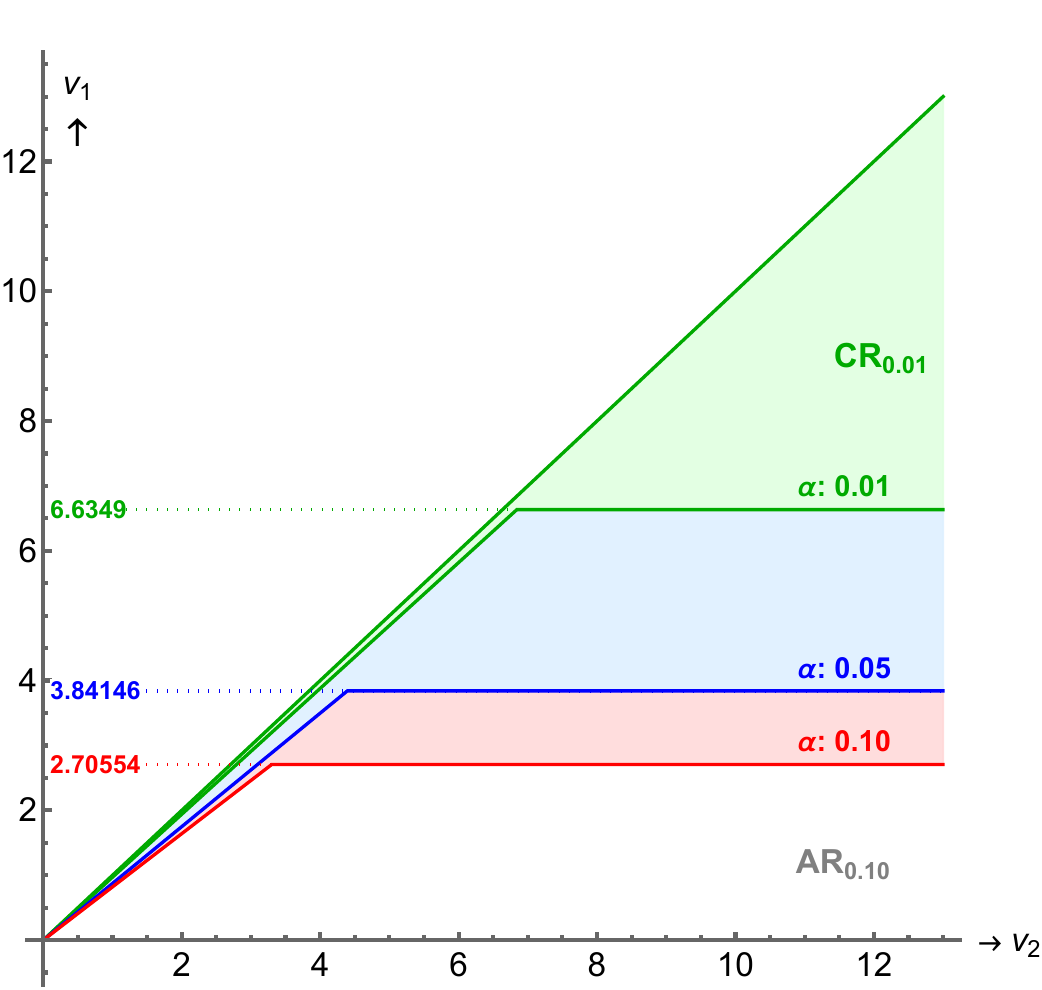}
\caption{CR Simply-augmented LR test} \label{fig:CRsimple} \end{subfigure}%
\end{tabular}%
\caption[short]{Critical regions for the two simple augmented LR tests.
Realizations in $CR_{0.05}\backslash CR_{0.10}$ of the truncated version
would reject at $.05$ but not at $.1$, showing incoherence. The
simply-augmented LR CRs are coherent subsets for decreasing $\protect\alpha $%
.}
\label{fig:CR_tr_augm_001_005_010_coher}
\end{figure}

The LR critical region has the desirable property that if an observed point $%
(v_{1},v_{2})$ falls in the rejection region at level $\alpha ,$ it also
falls in the rejection region at every level larger than $\alpha .$ That is,
the CR at a given level properly contains that at any smaller level. The LR
test is thus coherent for inference on $H_{0}.$ An important property of the
simply-augmented LR test as we have constructed it is that \textit{it
retains this coherency property}. This is perhaps best illustrated
graphically. Figure \ref{fig:CR_tr_augm_001_005_010_coher} shows the
respective critical regions for the levels $\alpha =.01,.05,$ and $.1$: for
the truncated version on the left, the simple version on the right. It is
clear that the proposed approach provides coherent inference on $H_{0}$ in
this sense, but only for the simply-augmented test, not the truncated
version. 

\subsection{p-values}

The coherence property just mentioned suggests that we can define a p-value
for any observed point $(v_{1},v_{2})$ by reference to the critical regions $%
CR_{b(\alpha )}.$ To do so, we simply determine the value of $\alpha ,$ say $%
\alpha _{0},$ for which the observed point lies on the boundary of the
critical region $CR_{b(\alpha _{0})}.$ Any point in the region $CR_{b(\alpha
_{0})}$ lies on a boundary with a smaller level than $\alpha _{0},$ and in
this sense are \textquotedblleft more extreme\textquotedblright\ under the
null hypothesis than the observed point. The value $\alpha _{0}$ then has a
natural interpretation as the p-value for the observed point.

To define $\alpha _{0}$ explicitly we make three observations: First, $%
b(\alpha )$ is strictly monotonic and therefore has an inverse, so for each $%
b\in \lbrack 0,1]$ there is a unique value $\alpha _{b}$ satisfying $%
b=b(\alpha _{b})$. \ Second, each $v_{1}\geq 0$ yields a value $\alpha
_{1}=1-G(v_{1})\in \lbrack 0,1].$ Third, every point $(v_{1},v_{2})\in V$
lies on either the horizontal part of the boundary of some $CR_{b(\alpha )},$
or on the sloping part. If it lies on the horizontal part, such that $%
v_{1}/v_{2}<b(\alpha _{1}),$ then $\alpha _{0}=\alpha _{1}=1-G(v_{1})$. If
it lies on the sloping part then $\alpha _{0}=b^{-1}(v_{1}/v_{2})$ and $%
v_{1}/v_{2}=b(\alpha _{0}).$

Determining the p-value is now straightforward using Table \ref{tab:alfa_b_z}%
: look up $v_{1}$ in the $\chi_{\alpha }^{2}$ column and note the
corresponding $\alpha _{1}.$ If $v_{1}/v_{2}\leq b(\alpha _{1})$ then this $%
\alpha _{1}$ is the p-value. If not, then $(v_{1},v_{2})$ must be on the
sloping part of the boundary, so look up $v_{1}/v_{2}$ in the $b(\alpha )$
column. The p-value is the corresponding $\alpha $. One can interpolate for
additional accuracy.

\section{Simulation and illustration}

\label{sec:Simulations}

In order to demonstrate the applicability of the test and performance more
broadly than the basic Gaussian setting, we carried out simulations using
nonnormal and heteroskedastic disturbances, as well as a logit as a
nonlinear model. Second, we include an empirical illustration from business.

\subsection{A simulation exercise}

\begin{figure}[h]
\begin{center}
\includegraphics[width=6.5in]{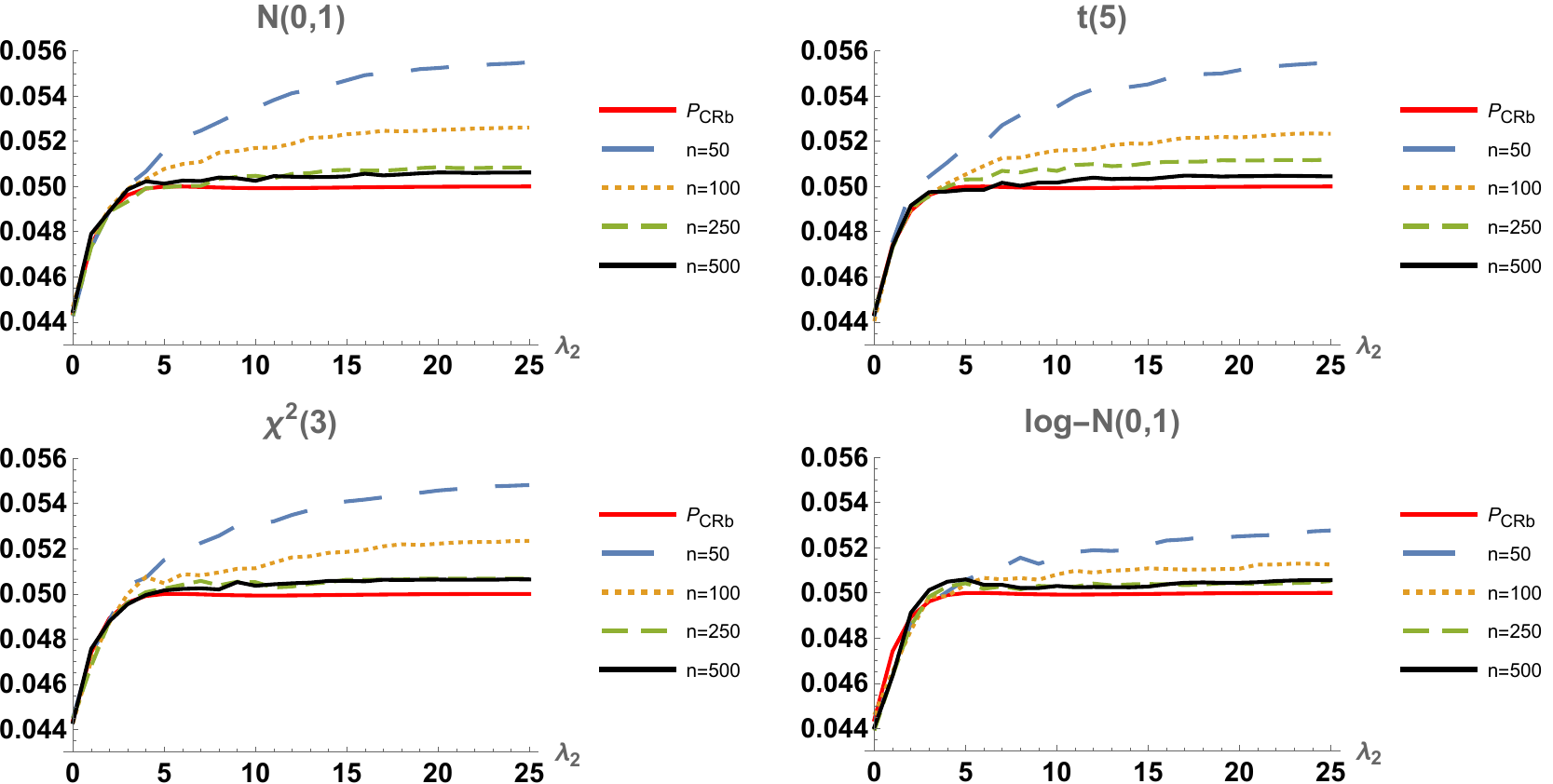}
\end{center}
\par
\vspace*{-6mm}
\caption{Finite-sample NRPs based on $10^{6}$ replications for different
distributions. The red line is not based on simulations but represents the
true NRP based on $P_{CR_{b}}(\protect\lambda _{2})$ that could be
interpreted as the NRP for $n=\infty $ without simulation error.}
\label{fig:MC_NonNormal}
\end{figure}

The asymptotic normal distribution of two independent $t$-statistics forms
the basis of the simply-augmented LR test. The approximation is valid for
various estimation methods and error distributions asymptotically, but may
be less accurate in small samples. Finite-sample critical values will
deviate from those of the normal distribution, even in the case of
homoskedastic normally distributed disturbances, when the two test
statistics $f_{1}$ and $f_{2}$ are $F$-distributed with differing degrees of
freedom, which further destroys the symmetry. To investigate the accuracy of
the approximation in different circumstances, we used simulations to
determine NRPs for different sample sizes and error distributions for $u_{1}$
and $u_{2}$: standard normal, student-$t$ (fat-tailed, with 5 degrees of
freedom), chi-squared (skewed, with 3 degrees of freedom), and log-normal
(skewed and fat-tailed), all standardized to mean zero and variance one.
Sample sizes of 50, 100, 250, and 500 were considered, and $x_{i}\sim N(0,1)$%
. Ordinary, rather than robust $t$-statistics are used. Given the
NRP-deviation from $0.05$ in Figure \ref{fig:NRP_LR_LRaug_05_10}, the
noncentrality parameter was chosen (approximately) as $\lambda _{2}\in
\{0,1,...,25\}$ by solving $\lambda _{2}=\theta _{2}^{2}/(n-2)$ in (A.4) of
Appendix A for $\theta _{2}$, and $\lambda _{1}=0$ implies $\theta _{1}=0$.
Figure \ref{fig:MC_NonNormal} shows the NRP based on $P_{CB_{b}}(\lambda
_{2})$, abbreviated to $P_{CB_{b}}$ in the figure's legend, and the
simulated NRPs based on $10^{6}$ replications. For the normally distributed
error case, the simulated NRPs are systematically higher than $P_{CB_{b}}$
with the largest deviation of approximately $0.006$ when $n=50$, in line
with an $F$-distribution based critical value being larger than based on the 
$\chi ^{2}$ distribution. The simulated NRPs quickly converge to $P_{CBb},$
however, as the sample size increases with almost no deviance for $n=500$.
In the case of errors that are standardized $t$-distributed or $\chi ^{2}$%
-distributed, the results are very similar, although the convergence to $%
P_{CB_{b}}$ is slower in the sample size. For the standardized log-normal
error distribution, the maximum deviation is larger/smaller than for the
other distributions for small/large values of the noncentrality parameter.
Overall, the simulation results confirm that the approximation made in (\ref%
{eq:asympdistr_t}) is very accurate, leading to a maximum overrejection of
only $0.0025$ when $n=100$ for a wide variety of error distributions.

We continue to investigate the efficacy of employing robust $t$-statistics
under homoskedasticity or in the presence of heteroskedasticity: $%
(u_{1,i},u_{2,i})^{\prime }\sim N(0,\sigma (x_{i})I_{2})$ with (i) $\sigma
(x_{i})=1$, (ii) $\sigma (x_{i})=|x_{i}|$, and (iii) $\sigma (x_{i})=\exp
(0.4x_{i})$. Table \ref{table:NRP_hetero100} shows that when ordinary $t$%
-statistics are applied, the NRP may be as high as $20.8\%$ in case (ii).
When robust $t$-statistics are used, the NRPs of the simply-augmented LR
test are within $[4.6\%,6.5\%]$, while those of the LR\ test can be as low
as $0.4\%$. The application of robust $t$-statistics only marginally
increases the NRPs under homoskedasticity. Hence, in empirical research, it
is recommended to use robust $t$-statistics. 
\spacingset{1.2}%
\begin{table}[t]
\caption{Simulated NRPs ($\times 100\%$) for various variance
specifications: (i) $\protect\sigma (x_{i})=1$, (ii) $\protect\sigma %
(x_{i})=|x_{i}|$, and (iii) $\protect\sigma (x_{i})=\exp (0.4x_{i})$: $H_{0}:%
\protect\theta _{1}\protect\theta _{2}=0$ (with $\protect\theta _{1}=0$), $%
n=100$, $10^{6}$ replications. Left panel employs ordinary $t$-statistics,
right panel utilizes the robust $t$-statistics. See also Appendix D for
additional sample sizes $n=50,250,500$ }
\label{table:NRP_hetero100}\vspace*{-3mm}
\par
\begin{center}
\begin{tabular}{lrrrrrrrrrrrrrrr}
& \multicolumn{7}{c}{Non-robust SE} &  & \multicolumn{7}{c}{Robust SE} \\ 
$n=100$ & \multicolumn{3}{c}{LR} &  & \multicolumn{3}{c}{$LR(b)$} &  & 
\multicolumn{3}{c}{LR} &  & \multicolumn{3}{c}{$LR(b)$} \\ 
$\theta _{2}$ & (i) & (ii) & (iii) &  & (i) & (ii) & (iii) &  & (i) & (ii) & 
(iii) &  & (i) & (ii) & (iii) \\ 
\cline{2-4}\cline{6-8}\cline{10-12}\cline{14-16}
0.0 & 0.3 & 4.1 & 1.5 &  & 4.6 & 7.6 & 5.5 &  & 0.4 & 0.5 & 0.5 &  & 4.6 & 
4.7 & 4.7 \\ 
0.14 & 1.5 & 7.6 & 3.8 &  & 5.0 & 10.5 & 7.0 &  & 1.7 & 1.3 & 1.5 &  & 5.1 & 
5.0 & 5.2 \\ 
0.39 & 5.0 & 18.4 & 10.4 &  & 5.2 & 19 & 10.9 &  & 5.6 & 4.8 & 5.5 &  & 5.7
& 6.0 & 6.3 \\ 
0.59 & 5.3 & 20.7 & 11.3 &  & 5.3 & 20.8 & 11.4 &  & 5.8 & 6.2 & 6.4 &  & 5.8
& 6.3 & 6.5 \\ \cline{2-4}\cline{6-8}\cline{10-12}\cline{14-16}
\end{tabular}%
\end{center}
\end{table}
\spacingset{1.9}%

\begin{figure}[h]
\begin{center}
\includegraphics[width=6.5in]{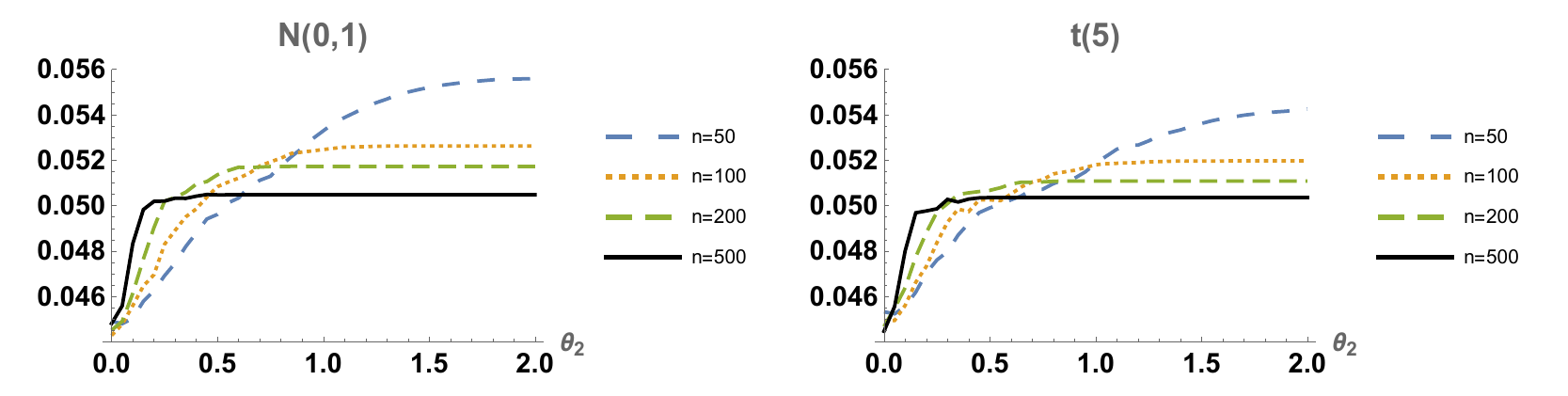}
\end{center}
\par
\vspace*{-6mm}
\caption{Finite-sample NRPs based on $10^{6}$ replications for $\protect%
\varepsilon _{i}\sim Logistic(0,1)$ in (\protect\ref{eq:yxmLogit}) and $%
u_{1}\sim N(0,1)$ or $u_{1}\sim t(5)$ in (\protect\ref{eq:mxLogit}).}
\label{fig:MC_Logit}
\end{figure}

The simply-augmented LR test is applicable as long as the $t$-ratios are
asymptotically standard normally distributed. This occurs more generally and
one example of a nonlinear model with binary dependent variable is the
following logit model with intercepts $\mu _{m}$ and $\mu _{y\ast }$: 
\begin{eqnarray}
m &=&\mu _{m}+\theta _{1}x+u_{1},  \label{eq:mxLogit} \\
y &=&\mathbbm{1}\{y^{\ast }>0\},\qquad y^{\ast }=\mu _{y\ast }+\tau x+\theta
_{2}m+\varepsilon ,  \label{eq:yxmLogit}
\end{eqnarray}%
where $\varepsilon _{i}\sim Logistic(0,1)$. The null of no mediation effect
is still $H_{0}:\theta _{1}\theta _{2}=0$, which can be tested using the
simply-augmented LR test, where $t_{1}$ and $t_{2}$ now denote the $t$%
-ratios of $\hat{\theta}_{1}$ and $\hat{\theta}_{2}$ in the estimated
regression (\ref{eq:mxLogit}) and logit model (\ref{eq:yxmLogit})
respectively. The simulation closely follows the specification used in %
\citet{MacKinnon2007}: $x$ is a dichotomous variable with equal numbers in
each group, $u_{1}$ is either standard normal, student-$t$ (fat-tailed, with
5 degrees of freedom), or chi-squared (skewed, with 3 degrees of freedom)
distributed, all standardized to mean zero and variance one. Parameter
values for $\theta _{2}\in \{0.0,0.05,...,2.0\}$ were chosen to encompass
the small (0.14), medium (0.39), large (0.59) and very large value (1.0)
considered by \citet{MacKinnon2007}, and $\theta _{1}=0$ (null hypothesis), $%
\tau =0$ (irrelevant by invariance), and $\mu _{m}=\mu _{y\ast }=0$. The
simulated NRPs in Figure \ref{fig:MC_Logit} are based on $10^{6}$
replications and are shown only for the normal and student-$t$ error
distributions since the results for the chi-squared distribution are almost
identical. The largest overrejection occurs in the normal case, with a
maximal deviation of $0.006$ when $n=50$. For sample sizes $n\geq 100$, the
new test is well-behaved with a maximum deviation of only $0.0026$. In the
student-$t$ case, the deviations from the nominal level seem to be even
smaller than in the normal case. For $n=500$ the figures are virtually
identical to the results in Figure \ref{fig:NRP_LR_LRaug_05_10} based on
asymptotic theory. In summary, the simply-augmented $LR$ test also performs
as expected in the important case of a binary dependent variable.

\subsection{Economic illustration}

To illustrate our simply-augmented LR test, we use data from 
\citet[Section
4.2]{hayes2017introduction}; this data set is called ESTRESS and can be
downloaded from \emph{www.afhayes.com}. The study involves entrepreneurs who
were members of a networking group for small business owners; see %
\citet{pollack2012moderating}. They answered an online survey about the
recent performance of their business and also about their emotional and
cognitive reactions to the economic climate. Hence, $Y$, $X$ and $M$ denote
disengagement from entrepreneurial activities ($withdraw$), economic stress (%
$estress$) and depressed affect ($affect$) respectively. There are also
three confounding variables $C_{1}$, $C_{2}$ and $C_{3}$ that are related to
entrepreneurial self-efficacy ($ese$), gender ($sex$) and length of time in
the business ($tenure$); see Figure 4 of \citet{hayes2017introduction} for a
causal diagram. We focus on females with short $tenure$ (less than 0.6
years). OLS\ gives the following results (showing ordinary $t$-values in
parentheses):\footnote{%
Misspecification tests do not reject normality or homoskedasticity, with the
following p-values:\newline
Jarque-Bera (normality): 0.5078 ($M$) \& 0.1848 ($Y$) and Breusch-Pagan
(homoskedasticity): 0.2016 ($M$) \& 0.1019 ($Y$).}%
\begin{eqnarray*}
\hat{M} &=&\underset{(2.037)}{2.0328}+\underset{(1.120)}{0.1606}X-\underset{%
(-1.802)}{0.2451}C_{1}+\underset{(0.705)}{0.4795}C_{3}, \\
\hat{Y} &=&\underset{(0.652)}{1.3229}-\underset{(-0.661)}{0.1776}X+\underset{%
(1.130)}{0.5249}M+\underset{(0.286)}{0.0772}C_{1}+\underset{(1.318)}{1.6398}%
C_{3}.
\end{eqnarray*}%
From these estimation results, we get the following test statistics: $%
(f_{1},f_{2})=(1.277,1.254)$ and $(v_{1},v_{2})=(1.254,1.277)$, so that $%
LR=1.254$. Since $LR<3.84$, the null of no mediation is not rejected at $.05$
level using the LR test or Sobel's test. Both $F$-statistics are very
similar, however, leading to a ratio $v_{1}/v_{2}=.982$ that is larger than $%
b(.05)=.8744$. The optimal simply-augmented test rejects, with a p-value of $%
.00593$ using interpolation and Table \ref{tab:alfa_b_z}. The new test
establishes a significant mediation effect that remains undetected by the
conservative LR\ and Sobel tests.

\section{Conclusion and closing comments}

\label{sec:conclusion}

We have demonstrated constructively that exact similar tests of the
no-mediation hypothesis exist for tests of the nominal levels that are
typically used in practice. Because the exact test described is incoherent,
we have also proposed a basic modification of the LR test which is called
the simply-augmented LR test. This new test to a large extent remedies the
two main deficiencies of the standard LR test: very small NRP for small
values of the nuisance parameter, and very poor power near the origin of the
parameter space. This simply-augmented test is extremely easy to understand
and apply in practice, and has the important property of coherence.

The earlier paper by \citet{VG2-2021}, in which the idea of augmenting the $%
LR$ critical region was first proposed for this problem, provides a more
sophisticated augmentation method. It performs somewhat better in terms of
NRP and power, but we show that it cannot be size correct.\ The test
proposed here has the advantage of extreme simplicity and ease of
implementation in empirical research. It requires only one additional number
in conjunction with the standard critical value. Moreover, it is coherent,
which is essential for deriving and interpreting p-values in applied work.
We give results for each percentile in the table at the end of the paper,
and show how p-values can be easily calculated and reported. A simulation
study confirms the asymptotic approximations with non-Gaussian disturbances,
with heteroskedastic disturbances, and in a relevant model for binary
dependent variables.

There is no doubt that both the approach discussed in this paper, along with
that of Van Garderen and Van Giersbergen, fall into the class of tests -
departures from the likelihood ratio method - that is frowned upon by %
\citet{Perlman1999}. Both lead to critical regions that imply rejection of
the null hypothesis when the observed sample point is close to the origin.
Moreover, the arguments claiming an improvement over the LR test are
certainly based firmly on the Neyman-Pearson criteria of size and power. If
one does not approve, the LR test is still available of course.\clearpage

\spacingset{1.6}%

\clearpage\pagebreak

\spacingset{1.2}%
\begin{table}[t!]
\caption{Optimal $b(\protect\alpha )$ and $\protect\chi _{\protect\alpha%
}^{2} $ for percentiles $\protect\alpha \in [0,1]$ for the simply-augmented
test. \newline
Reject $H_{0}$: no mediation at level $\protect\alpha $ if $v_{1}\geq 
\protect\chi _{\protect\alpha }^{2}$ and/or $v_{1}/v_{2}\geq b(\protect\alpha%
)$.\newline
P-value: \texttt{look up} $v_{1}$ in $\protect\chi _{\protect\alpha }^{2}$
column and note corresponding $\protect\alpha$ (possibly using linear
interpolation); \texttt{if} $v_{1}/v_{2}\leq b(\protect\alpha)$ \texttt{then}
p-value $=\protect\alpha$ \texttt{else look up} $v_{1}/v_{2}$ in $b(\protect%
\alpha)$ column; p-value is corresponding $\protect\alpha %
(=b^{-1}(v_{1}/v_{2}))$.}
\label{tab:alfa_b_z}\vspace*{-6mm}
\par
\begin{center}
{\footnotesize \ 
\begin{tabular}{|c|c|c|c|c|c|c|c|c|c|c|}
\hline
$\alpha $ & $b(\alpha )$ & $\chi_{\alpha }^{2}$ &  & $\alpha $ & $b(\alpha ) 
$ & $\chi_{\alpha }^{2}$ &  & $\alpha $ & $b(\alpha )$ & $\chi_{\alpha }^{2}$
\\ \hline
0.00 & 1.0000000 & $\infty$ &  & 0.34 & 0.61042 & 0.9104313 &  & 0.68 & 
0.25136 & 0.1701258 \\ 
0.01 & 0.9696632 & 6.6348966 &  & 0.35 & 0.60140 & 0.8734571 &  & 0.69 & 
0.23958 & 0.1590854 \\ 
0.02 & 0.9418969 & 5.4118944 &  & 0.36 & 0.59230 & 0.8378932 &  & 0.70 & 
0.22782 & 0.1484719 \\ 
0.03 & 0.9168391 & 4.7092922 &  & 0.37 & 0.58312 & 0.8036645 &  & 0.71 & 
0.21612 & 0.1382770 \\ 
0.04 & 0.8943890 & 4.2178846 &  & 0.38 & 0.57384 & 0.7707019 &  & 0.72 & 
0.20446 & 0.1284927 \\ 
0.05 & 0.8744040 & 3.8414588 &  & 0.39 & 0.56448 & 0.7389420 &  & 0.73 & 
0.19288 & 0.1191116 \\ 
0.06 & 0.8568159 & 3.5373846 &  & 0.40 & 0.55502 & 0.7083263 &  & 0.74 & 
0.18138 & 0.1101266 \\ 
0.07 & 0.8445200 & 3.2830203 &  & 0.41 & 0.54548 & 0.6788007 &  & 0.75 & 
0.17002 & 0.1015310 \\ 
0.08 & 0.8345800 & 3.0649017 &  & 0.42 & 0.53582 & 0.6503152 &  & 0.76 & 
0.15878 & 0.0933185 \\ 
0.09 & 0.8250200 & 2.8743734 &  & 0.43 & 0.52608 & 0.6228235 &  & 0.77 & 
0.14772 & 0.0854831 \\ 
0.10 & 0.8157800 & 2.7055435 &  & 0.44 & 0.51624 & 0.5962824 &  & 0.78 & 
0.13684 & 0.0780191 \\ 
0.11 & 0.8067600 & 2.5542213 &  & 0.45 & 0.50628 & 0.5706519 &  & 0.79 & 
0.12618 & 0.0709213 \\ 
0.12 & 0.7979200 & 2.4173209 &  & 0.46 & 0.49624 & 0.5458947 &  & 0.80 & 
0.11576 & 0.0641848 \\ 
0.13 & 0.7892200 & 2.2925045 &  & 0.47 & 0.48608 & 0.5219760 &  & 0.81 & 
0.10560 & 0.0578047 \\ 
0.14 & 0.7806400 & 2.1779592 &  & 0.48 & 0.47582 & 0.4988633 &  & 0.82 & 
0.09576 & 0.0517767 \\ 
0.15 & 0.7721400 & 2.0722509 &  & 0.49 & 0.46544 & 0.4765263 &  & 0.83 & 
0.08624 & 0.0460968 \\ 
0.16 & 0.7637000 & 1.9742261 &  & 0.50 & 0.45498 & 0.4549364 &  & 0.84 & 
0.07706 & 0.0407610 \\ 
0.17 & 0.7553200 & 1.8829433 &  & 0.51 & 0.44440 & 0.4340671 &  & 0.85 & 
0.06828 & 0.0357658 \\ 
0.18 & 0.7469600 & 1.7976241 &  & 0.52 & 0.43372 & 0.4138933 &  & 0.86 & 
0.05992 & 0.0311078 \\ 
0.19 & 0.7386200 & 1.7176176 &  & 0.53 & 0.42294 & 0.3943916 &  & 0.87 & 
0.05202 & 0.0267841 \\ 
0.20 & 0.7303000 & 1.6423744 &  & 0.54 & 0.41206 & 0.3755398 &  & 0.88 & 
0.04458 & 0.0227917 \\ 
0.21 & 0.7219800 & 1.5714263 &  & 0.55 & 0.40108 & 0.3573172 &  & 0.89 & 
0.03762 & 0.0191281 \\ 
0.22 & 0.7136400 & 1.5043712 &  & 0.56 & 0.39000 & 0.3397042 &  & 0.90 & 
0.03122 & 0.0157908 \\ 
0.23 & 0.7052800 & 1.4408614 &  & 0.57 & 0.37882 & 0.3226825 &  & 0.91 & 
0.02534 & 0.0127777 \\ 
0.24 & 0.6969000 & 1.3805940 &  & 0.58 & 0.36756 & 0.3062346 &  & 0.92 & 
0.02006 & 0.0100869 \\ 
0.25 & 0.6885000 & 1.3233037 &  & 0.59 & 0.35620 & 0.2903443 &  & 0.93 & 
0.01536 & 0.0077167 \\ 
0.26 & 0.6800400 & 1.2687570 &  & 0.60 & 0.34478 & 0.2749959 &  & 0.94 & 
0.01126 & 0.0056656 \\ 
0.27 & 0.6715400 & 1.2167470 &  & 0.61 & 0.33328 & 0.2601749 &  & 0.95 & 
0.00780 & 0.0039321 \\ 
0.28 & 0.6630000 & 1.1670899 &  & 0.62 & 0.32170 & 0.2458676 &  & 0.96 & 
0.00496 & 0.0025154 \\ 
0.29 & 0.6544000 & 1.1196214 &  & 0.63 & 0.31008 & 0.2320608 &  & 0.97 & 
0.00276 & 0.0014144 \\ 
0.30 & 0.6457400 & 1.0741942 &  & 0.64 & 0.29840 & 0.2187422 &  & 0.98 & 
0.00122 & 0.0006285 \\ 
0.31 & 0.6370000 & 1.0306758 &  & 0.65 & 0.28666 & 0.2059001 &  & 0.99 & 
0.00030 & 0.0001571 \\ 
0.32 & 0.6282200 & 0.9889465 &  & 0.66 & 0.27492 & 0.1935236 &  & 1.00 & 
0.00000 & 0.0000000 \\ 
0.33 & 0.6193600 & 0.9488978 &  & 0.67 & 0.26314 & 0.1816021 &  &  &  &  \\ 
\hline
\end{tabular}
}
\end{center}
\end{table}
\clearpage\pagebreak

\renewcommand{\thetheorem}{\Alph{section}.\arabic{theorem}}
\renewcommand{\thelemma}{\Alph{section}.\arabic{lemma}}
\renewcommand{\theproposition}{\Alph{section}.\arabic{proposition}}
\renewcommand{\thecorollary}{\Alph{section}.\arabic{corollary}}
\renewcommand{\theassumption}{\Alph{section}.\arabic{assumption}}
\renewcommand{\thesection}{\Alph{section}}
\renewcommand{\theequation}{\Alph{section}.\arabic{equation}}
\renewcommand{\thefigure}{\Alph{section}.\arabic{figure}}
\renewcommand{\thetable}{\Alph{section}.\arabic{table}}
\numberwithin{equation}{section}
\setcounter{section}{0}
\setcounter{table}{0}
\setcounter{figure}{0}

\spacingset{1.9}%

{\centering \LARGE\textbf{Appendix: Improved Tests for Mediation}}

\section{Theory and proofs}

\label{sec:appendix_a_proofs}

\subsection{Invariance under Gaussianity}

The testing problem for $H_{0}:\theta _{1}\theta _{2}=0,$ respects a number
of symmetries. It allows restricting our attention for an optimal solution
in terms of the maximal invariant statistic, see e.g. 
\citet[Section
5.3]{Davison2003}, which we derive now under a Gaussianity assumption:%
\begin{eqnarray}
m|x\ \ \ &\sim &N(x\theta _{1},\sigma _{11}I_{n}),  \label{eq:Gauss_m_x} \\
y|x,m &\sim &N(x\tau +\theta _{2}m,\sigma _{22}I_{n}).  \label{eq:Gauss_y_xm}
\end{eqnarray}

The statistics of interest are the sufficient statistics in the Gaussian
model, which are simply the OLS estimates for (\ref{eq:Gauss_m_x}) and (\ref%
{eq:Gauss_y_xm}), including the residual sums of squares, i.e. $(\hat{\theta}%
_{1},s_{11},\hat{\tau},\hat{\theta}_{2},s_{22})$ as defined in the proof of
Theorem \ref{Th:invariancetstat}. This theorem derives the maximal
invariants which reduce the five-dimensional sufficient statistic to just
two dimensions.

\begin{theorem}
\label{Th:invariancetstat}In the model (\ref{eq:Gauss_m_x}) and (\ref%
{eq:Gauss_y_xm}), the problem of testing $H_{0}:\theta _{1}\theta _{2}=0$ is
invariant under the group $\mathbf{K}=\{a_{1},a_{2},c:a_{1},a_{2}>0,c\in 
\mathbb{R}\}$ of transformations acting on $(\hat{\theta}_{1},s_{11},\hat{%
\tau},\hat{\theta}_{2},s_{22})$ by 
\begin{equation}
(\hat{\theta}_{1},s_{11},\hat{\tau},\hat{\theta}_{2},s_{22})\mapsto (\sqrt{%
a_{1}}\hat{\theta}_{1},a_{1}s_{11},\sqrt{a_{2}}\left( \hat{\tau}+c\right) ,%
\sqrt{a_{2}/a_{1}}\hat{\theta}_{2},a_{2}s_{22}).  \notag
\end{equation}%
A sample-space maximal invariant under this group of transformations is 
\begin{equation}
T_{1}=\hat{\theta}_{1}/\sqrt{\tfrac{1}{n-1}s_{11}/s_{xx}},\ T_{2}=\hat{\theta%
}_{2}/\sqrt{\tfrac{1}{n-2}s_{22}/s_{11}},
\end{equation}%
The induced group of transformations on $(\theta _{1},\sigma _{11},\tau
,\theta _{2},\sigma _{22})$ is defined by 
\begin{equation*}
(\theta _{1},\sigma _{11},\tau ,\theta _{2},\sigma _{22})\mapsto (\sqrt{a_{1}%
}\theta _{1},a_{1}\sigma _{11},\sqrt{a_{2}}(\tau +c),\sqrt{a_{2}/a_{1}}%
\theta _{2},a_{2}\sigma _{22}).
\end{equation*}%
A parameter-space maximal invariant under the induced group is: 
\begin{equation}
\mu _{1}=\theta _{1}/\sqrt{\sigma _{11}/s_{xx}},\ \mu _{2}=\theta _{2}/\sqrt{%
\tfrac{1}{n-2}\sigma _{22}/\sigma _{11}}.  \label{eq:mus}
\end{equation}%
The distribution of $(T_{1},T_{2})$ depends only on $(\mu _{1},\mu _{2}).$
\end{theorem}

\textbf{Proof. }The maximal invariants $T_{1},T_{2}$ are the usual $t$%
-statistics for testing $\theta _{1}=0$ and$\ \theta _{2}=0$. Under the
Gaussianity assumption in equations (\ref{eq:Gauss_m_x}) and (\ref%
{eq:Gauss_y_xm}), the MLEs for $\theta _{1}$ and $(\tau ,\theta _{2})$ are
the OLS estimators and the MLEs for variances $\sigma _{11}$ and $\sigma
_{22}$ are the residual sums of squares $s_{11}$ and $s_{22}$ divided by
sample size $n$:%
\begin{eqnarray*}
\hat{\theta}_{1} &=&(x^{\prime }x)^{-1}x^{\prime }m,\ \ \ \ s_{11}=m^{\prime
}M_{x}m \\
\binom{\hat{\tau}}{\hat{\theta}_{2}} &=&[(x,m)^{\prime
}(x,m)]^{-1}(x,m)^{\prime }y,\ \ \ \ s_{22}=y^{\prime }M_{x,m}y.
\end{eqnarray*}%
Here, for any matrix $A$ of full column rank, $M_{A}=I_{n}-A(A^{\prime
}A)^{-1}A^{\prime }$. The distributions of the sufficient statistics are,
respectively:%
\begin{equation*}
\binom{\hat{\tau}}{\hat{\theta}_{2}}|m\sim N\left( \binom{\tau }{\theta _{2}}%
,\sigma _{22}[(x,m)^{\prime }(x,m)]^{-1}\right) ,
\end{equation*}%
$s_{11}/\sigma _{11}\sim \chi ^{2}(n-1),$ $\hat{\theta}_{1}\sim N(\theta
_{1},\sigma _{11}/s_{xx}),$ where $s_{xx}=x^{\prime }x,$ and $s_{22}/\sigma
_{22}\sim \chi ^{2}(n-2)$. The joint density of the sufficient statistics
under Gaussian assumptions may be written down directly from these facts,
and is equivalent to the likelihood for $(\theta _{1},\sigma _{11},\tau
,\theta _{2},\sigma _{22}).$ The joint distribution of the sufficient
statistics is a product of the form: 
\begin{eqnarray*}
f(\hat{\theta}_{1},s_{11},\hat{\tau},\hat{\theta}_{2},s_{22}) &=&N(\theta
_{1},\sigma _{11}(x^{\prime }x)^{-1})\times \sigma _{11}\chi
_{n-1}^{2}\times N\left( \tau -\hat{\theta}_{1}\left( \hat{\theta}%
_{2}-\theta _{2}\right) ,\sigma _{11}\left( x^{\prime }x\right) ^{-1}\right) 
\\
&&~~~\times N\left( \theta _{2},\sigma _{22}/s_{11}\right) \times \left(
\sigma _{22}\chi _{n-2}^{2}\right) .
\end{eqnarray*}

The transformations $s_{11}\mapsto a_{1}s_{11}$ and $s_{22}\mapsto
a_{2}s_{22}$ with $a_{1},a_{2}>0$ leave the joint density of $\left(
s_{11},s_{22}\right) $ in the same family with $\left( \sigma _{11},\sigma
_{22}\right) $ replaced by $\left( a_{1}\sigma _{11},a_{2}\sigma
_{22}\right) $ and have no bearing on the hypothesis under test. The same
parameters $\sigma _{11}$ and $\sigma _{22}$ are present in the other
components, so we need to transform the remaining variables accordingly,
namely by: $\hat{\theta}_{1}\mapsto \sqrt{a_{1}}\hat{\theta}_{1}$ and $\hat{%
\theta}_{2}\mapsto \sqrt{a_{2}/a_{1}}\hat{\theta}_{2}$. And, since $\tau $
is not involved in the inference problem, we may transform $\hat{\tau}$ by
the affine transformation $\hat{\tau}\mapsto \sqrt{a_{2}}\left( \hat{\tau}%
+c\right) $.

These transformations preserve the family of distributions for the
sufficient statistics (and MLEs), and the induced transformation on the
mediation effect is that $\theta _{1}\theta _{2}\mapsto \sqrt{a_{2}}\theta
_{1}\theta _{2}$. Thus, the transformations do not change the truth or
falsity of the hypothesis under test (i.e. $H_{0}$ is true before iff it is
true after the transformation). The transformations on $\hat{\tau}$ are
transitive, so no invariant test can depend on $\hat{\tau}.$ We can
therefore restrict attention to the four remaining statistics\textbf{\ }$%
\left( \hat{\theta}_{1},s_{11},\hat{\theta}_{2},s_{22}\right) $, and the
group $\mathbf{K}$, say, of (scale) transformations of them. The invariance
of $(T_{1},T_{2})$ under the transformations is obvious. To show that $%
(T_{1},T_{2})$ are maximal we need to show that $T_{1}(\hat{\theta}_{1},\hat{%
\theta}_{2},s_{11},s_{22})=T_{1}(\tilde{\theta}_{1},\tilde{\theta}_{2},%
\tilde{s}_{11},\tilde{s}_{22})$ and $T_{2}(\hat{\theta}_{1},\hat{\theta}%
_{2},s_{11},s_{22})=T_{2}(\tilde{\theta}_{1},\tilde{\theta}_{2},\tilde{s}%
_{11},\tilde{s}_{22})$ implies that there exists a group element $K\in 
\mathbf{K}$ such that $(\tilde{\theta}_{1},\tilde{\theta}_{2},\tilde{s}_{11},%
\tilde{s}_{22})=K(\hat{\theta}_{1},\hat{\theta}_{2},s_{11},s_{22})$.

Thus, assume that 
\begin{equation*}
\hat{\theta}_{1}/\sqrt{s_{11}/s_{xx}}=\tilde{\theta}_{1}/\sqrt{\tilde{s}%
_{11}/s_{xx}}
\end{equation*}%
and 
\begin{equation*}
\hat{\theta}_{2}/\sqrt{\tfrac{1}{n-2}s_{22}/s_{11}}=\tilde{\theta}_{2}/\sqrt{%
\tfrac{1}{n-2}\tilde{s}_{22}/\tilde{s}_{11}}.
\end{equation*}%
Then $\tilde{\theta}_{1}=\sqrt{a_{1}}\hat{\theta}_{1}$ with $a_{1}=\tilde{s}%
_{11}/s_{11}$, and $\tilde{\theta}_{2}=\sqrt{a_{2}/a_{1}}\hat{\theta}_{2}$
with $a_{2}=\tilde{s}_{22}/s_{22}$. Since also $\tilde{s}_{11}=a_{1}s_{11}$,
and $\tilde{s}_{22}=a_{2}s_{22}$, this shows that the invariance of $%
(T_{1},T_{2})$ implies that the two sets of statistics are related by a
group element, so $(T_{1},T_{2})$ are indeed maximal. The same argument
applies to the induced group acting on the parameter space, and the last
statement is a well-known property of maximal invariants.

\bigskip

\textbf{Proposition 1} can be proved using the assumptions and theorems in %
\citet{White1980} as is explicitly done in \citet[Proposition 1]{VG2-2021}.

\subsection{Distributions of the order statistics}

The noncentral $\chi _{\kappa }^{2}$ density, $g_{\kappa }(f;\lambda ),$
with noncentrality $\lambda $, plays a central role throughout this paper
and its appendix.\ It can be expressed in several ways, including the
Poisson mixture exploited in the proof of Proposition 5: 
\begin{equation*}
g_{\kappa }(f;\lambda )=\exp \left\{ -\frac{1}{2}\lambda \right\}
\sum_{j=0}^{\infty }\frac{(\lambda /2)^{j}}{j!}g_{\kappa +2j}(f),
\end{equation*}%
where $g_{\kappa }(f)=[2^{\frac{\kappa }{2}}\Gamma (\frac{\kappa }{2}%
)]^{-1}\exp \{-\frac{1}{2}f\}f^{\frac{\kappa }{2}-1}$ denotes the $\chi
_{\kappa }^{2}$ density function, and we write $g_{1}(f)$ simply as $g(f).$
The corresponding CDFs are denoted by $G_{\kappa }(\cdot ;\lambda ),$ and $%
G_{\kappa }(\cdot )$ in the central case, the subscript being omitted when $%
\kappa =1.$ For $\alpha \in \lbrack 0,1],$ we define $\chi _{\alpha }^{2}$
by $G(\chi _{\alpha }^{2})=1-\alpha .$

From Equation (6) in \citet{Vaughan1972}, the joint density of the order
statistics for $(v_{1},v_{2})\in V=\{(v_{1},v_{2});0\leq v_{1}\leq
v_{2}<\infty \}$ is as given in part (i) of the following proposition, which
also gives complete details of the distribution of the order statistics$:%
\footnote{%
Since the order statistics are maximal invariants under the action of the
symmetric group $S_{2}$ on $(f_{1},f_{2}),$ the main result in part (i) can
also be obtained by invoking Stein's method of obtaining the density of the
maximal invariant by averaging the joint density over the group.}$

\begin{proposition}
\label{prop:distr_max_invariant} Let $g(v;\lambda )$ be the noncentral $\chi
_{1}^{2}$ density with noncentrality parameter $\lambda .$\newline
(i) The joint density of the order statistics on the region $V$ is given by 
\begin{equation}
pdf(v_{1},v_{2}|\lambda _{1},\lambda _{2})=\left[ g(v_{1};\lambda
_{1})g(v_{2};\lambda _{2})+g(v_{2};\lambda _{1})g(v_{1};\lambda _{2})\right]
.  \label{joint}
\end{equation}

(ii) When either $\lambda _{1}=0$ or $\lambda _{2}=0$ the null density, with 
$\lambda =\max \{\lambda _{1},\lambda _{2}\}$, is 
\begin{equation}
pdf(v_{1},v_{2}|\lambda )=g(v_{1})g(v_{2};\lambda )+g(v_{2})g(v_{1};\lambda
);  \label{nullden}
\end{equation}

(iii) The marginal density of the smaller order statistic $v_{1},$ i.e., the 
$LR$ statistic, is%
\begin{equation}
pdf(v_{1}|\lambda _{1},\lambda _{2})=g(v_{1};\lambda _{1})\left[
1-G(v_{1};\lambda _{2})\right] +g(v_{1};\lambda _{2})\left[
1-G(v_{1};\lambda _{1})\right] ,v_{1}\geq 0,  \label{eq:marg_dens_v1}
\end{equation}%
with corresponding CDF%
\begin{equation*}
H(v_{1};\lambda _{1},\lambda _{2})=G(v_{1};\lambda _{1})+G(v_{1};\lambda
_{2})-G(v_{1};\lambda _{1})G(v_{1};\lambda _{2}).
\end{equation*}
\end{proposition}

\noindent \textbf{Proof of Proposition \ref{prop:distr_max_invariant} and
remarks}

Part (i) is direct from \citet{Vaughan1972}. Part (iii) is simply the fact
that, on integrating over $v_{1}<v_{2}<\infty ,$ we have 
\begin{equation*}
pdf(v_{1}|\lambda _{1},\lambda _{2})=g(v_{1};\lambda
_{1})\int_{v>v_{1}}g(v;\lambda _{2})dv+g(v_{1};\lambda
_{2})\int_{v>v_{1}}g(v;\lambda _{1})dv.
\end{equation*}

\emph{\noindent Remarks}

(i) We can specialize these results for the null case when one noncentrality
parameter vanishes 
\begin{eqnarray*}
pdf(v_{1}|\lambda ) &=&g(v_{1})[1-G(v_{1};\lambda )]+g(v_{1};\lambda
)[1-G(v_{1})],~ \\
H(v;\lambda ) &=&G(v)+G(v;\lambda )-G(v)G(v;\lambda ), \\
1-H(v;0) &=&1-2G(v)+G(v)^{2}=(1-G(v))^{2},
\end{eqnarray*}%
where $\lambda $ is the nonzero noncentrality parameter.

(ii) It is trivial to check that the derivatives of $H(v;\lambda )$ and $%
H(v_{1};\lambda _{1},\lambda _{2})$ yield the densities given in (\ref%
{nullden}) and (\ref{eq:marg_dens_v1}).

(iii) Observe that 
\begin{equation*}
(\partial /\partial v_{1})[G(v_{1};\lambda _{1})G(v_{1};\lambda
_{2})]=[g(v_{1};\lambda _{1})G(v_{1};\lambda _{2})+g(v_{1};\lambda
_{2})G(v_{1};\lambda _{1})].
\end{equation*}%
This and similar identities are useful, for example, to verify that the
joint density integrates to one:%
\begin{eqnarray*}
\int_{v_{2}>0}\int_{0<v_{1}<v_{2}}pdf(v_{1},v_{2}|\lambda _{1},\lambda
_{2})dv_{1}dv_{2} &=&\int_{v_{2}>0}\left[ G(v_{2};\lambda
_{1})g(v_{2};\lambda _{2})+G(v_{2};\lambda _{2})g(v_{2};\lambda _{1})\right]
dv_{2} \\
&=&\int_{v_{2}>0}(\partial /\partial v_{2})[G(v_{2};\lambda
_{1})G(v_{2};\lambda _{2})]dv_{2} \\
&=&[G(v_{2};\lambda _{1})G(v_{2};\lambda _{2})]_{0}^{\infty }=1.
\end{eqnarray*}

It is not difficult to obtain the following probabilities under the null:
for any $z>0$,%
\begin{eqnarray*}
P_{A_{1}}(\lambda ;z) &=&[1-G(z)][1-G(z;\lambda )], \\
P_{A_{2}}(\lambda ;z) &=&G(z)[1-G(z;\lambda )]+G(z;\lambda )[1-G(z)], \\
P_{A_{3}}(\lambda ;z) &=&G(z)G(z;\lambda ),\text{ so that} \\
P_{A_{2}\cup A_{3}}(\lambda ;z) &=&G(z)+G(z;\lambda )-G(z)G(z;\lambda ).
\end{eqnarray*}

\subsection{Proof of Theorem 1}

Excluding the region $A_{r}(z)$ from $A_{3}$ leaves $r+1$ disjoint triangles
lying along the $45^{\circ }$ line, the region $w_{r}(z)\subset A_{3},$ and
it is easy to see that the null probability content of this region is 
\begin{equation*}
P_{w_{r}(z)}(\lambda )=(1-\alpha -G(z_{r}))G(\chi _{\alpha }^{2};\lambda
)-\sum_{i=1}^{r}\left[ G(z_{i+1})-2G(z_{i})+G(z_{i-1})\right]
G(z_{i};\lambda ).
\end{equation*}%
This differs from the target value $\alpha G(\chi _{\alpha }^{2};\lambda )$
by 
\begin{equation*}
(1-2\alpha -G(z_{r}))G(\chi _{\alpha }^{2};\lambda )-\sum_{i=1}^{r}\left[
G(z_{i+1})-2G(z_{i})+G(z_{i-1})\right] G(z_{i};\lambda ).
\end{equation*}%
This is a linear combination of $r+1$ noncentral chi-square CDFs, the $%
G(z_{i};\lambda )$ for $i\in \{1,...,r\}$, and $G(z_{r+1};\lambda )=G(\chi
_{\alpha }^{2};\lambda ),$ and vanishes for all $\lambda $ if and only if
the coefficients of all $r+1$ terms that involve $\lambda $ vanish. With $%
\alpha =1-G(z_{r+1}),$ these conditions give rise to a system of $r+1$
linear equations in $r+1$ unknowns $G(z_{i}),i=1,...,r+1$ (we take $z_{0}=0,$
so that $G(z_{0})=0),$ and these determine $z_{1},...,z_{r}$ and $%
z_{r+1}=\chi _{\alpha }^{2},$ hence $\alpha $. \ It is easy to see that the
matrix of the system is non-singular, so the solution is unique, and it is
straightforward to check that the solution is: 
\begin{equation*}
G(z_{i})=\frac{i}{r+2},i=1,...,r+1,
\end{equation*}%
so that $\alpha =1-G(z_{r+1})=(r+2)^{-1}.$

\subsection{Proof of Theorem 2}

Define a rectangle $R$ inside $AR_{LR}$ by first defining $\chi _{\alpha
+\epsilon }^{2}=z_{1}$ and $\chi _{\alpha -\epsilon }^{2}=z_{2}$ for $%
0<\epsilon \leq \alpha \leq \frac{1}{2},$ such that $1-G\left( z_{1}\right)
=\alpha +\epsilon ,$ $1-G\left( z_{2}\right) =\alpha -\epsilon $, $G(\chi
_{\alpha }^{2})-G(z_{1})=\epsilon ,$ $G(z_{2})-G(\chi _{\alpha
}^{2})=\epsilon ,$ $z_{1}\leq \chi _{\alpha }^{2}\leq z_{2},$ and then%
\begin{equation*}
R=\{v\in 
\mathbb{R}
^{2}\mid z_{1}<v_{1}<\chi _{\alpha }^{2},\chi _{\alpha }^{2}<v_{2}<z_{2}\},
\end{equation*}%
in the top left-hand corner of $A_{2}$. The probability content of $R$ is: 
\begin{eqnarray*}
\Pr \left[ R\right]  &=&\int_{\chi _{\alpha }^{2}}^{z_{2}}\int_{z_{1}}^{\chi
_{\alpha }^{2}}g(v_{1})g(v_{2};\lambda )+g(v_{2})g(v_{1};\lambda
)dv_{1}dv_{2} \\
&=&\int_{\chi _{\alpha }^{2}}^{z_{2}}\left[ G(\chi _{\alpha }^{2})-G\left(
z_{1}\right) \right] \cdot g(v_{2};\lambda )+g(v_{2})\cdot \left[ G(\chi
_{\alpha }^{2};\lambda )-G(z_{1};\lambda )\right] dv_{2} \\
&=&\left[ G(\chi _{\alpha }^{2})-G\left( z_{1}\right) \right] \cdot \left[
G(z_{2};\lambda )-G(\chi _{\alpha }^{2};\lambda )\right] +\left[
G(z_{2})-G(z_{\alpha })\right] \cdot \left[ G(\chi _{\alpha }^{2};\lambda
)-G(z_{1};\lambda )\right]  \\
&=&\epsilon \cdot \left[ G(z_{2};\lambda )-G(\chi _{\alpha }^{2};\lambda )%
\right] +\epsilon \cdot \left[ G(\chi _{\alpha }^{2};\lambda
)-G(z_{1};\lambda )\right]  \\
&=&\epsilon \cdot \left[ G(z_{2};\lambda )-G(z_{1};\lambda )\right] .
\end{eqnarray*}%
The rejection probability of the LR\ test augmented with $R$ is: 
\begin{equation*}
\Pr \left[ CR_{LR}\right] +\Pr \left[ R\right] =\alpha \cdot \left[ 1-G(\chi
_{\alpha }^{2};\lambda )\right] +\epsilon \cdot \left[ G(z_{2};\lambda
)-G(z_{1};\lambda )\right] .
\end{equation*}%
Hence the test is correctly sized iff%
\begin{equation*}
-\alpha \cdot G(\chi _{\alpha }^{2};\lambda )+\epsilon \cdot \left[
G(z_{2};\lambda )-G(z_{1};\lambda )\right] \leq 0\ \ \forall \lambda \geq 0
\end{equation*}%
or%
\begin{equation}
\frac{G(z_{2};\lambda )-G(z_{1};\lambda )}{G(\chi _{\alpha }^{2};\lambda )}%
\leq \frac{\alpha }{\epsilon }\ \ \forall \lambda \geq 0.  \label{eq:Ratio}
\end{equation}%
We now prove that this inequality cannot hold by constructing a lower-bound
and showing that it exceeds $\frac{\alpha }{\epsilon }$ for any $\alpha $
and $\epsilon $, as $\lambda \rightarrow \infty .$ The bound is constructed
by deriving a lower bound for the numerator and an upper bound for the
denominator. First note that the CDF\ of the noncentral Chi-square
distribution with one degree of freedom can be written as\footnote{%
By a simple transformation of the random variable $Z=X^{2}$ with $X\sim N(%
\sqrt{\lambda },1)$.}%
\begin{eqnarray}
G\left( z;\lambda \right)  &=&\Phi (\sqrt{z}-\sqrt{\lambda })+\Phi (\sqrt{z}+%
\sqrt{\lambda })-1  \notag \\
&=&\Phi (\sqrt{\lambda }+\sqrt{z})-\Phi (\sqrt{\lambda }-\sqrt{z}),
\label{eq:AlternativeG}
\end{eqnarray}%
where $\Phi (\cdot )$ denotes the cdf of the standard normal $N(0,1)$ and
the second line is written such that the argument of $\Phi (\cdot )$ is
positive when $\lambda >z$. The numerator in (\ref{eq:Ratio}) can therefore
be written as%
\begin{equation}
G(z_{2};\lambda )-G(z_{1};\lambda )=[\Phi (\sqrt{\lambda }-\sqrt{z_{1}}%
)-\Phi (\sqrt{\lambda }-\sqrt{z_{2}})]+[\Phi (\sqrt{\lambda }+\sqrt{z_{2}}%
)-\Phi (\sqrt{\lambda }+\sqrt{z_{1}})].  \label{eq:LowerNumerator4terms}
\end{equation}%
Using the Hermite-Hadamard inequality for convex $f\left( x\right) ,$ which
states:%
\begin{equation}
f\left( \frac{a+b}{2}\right) \leq \frac{1}{b-a}\int_{a}^{b}f(x)~dx\leq \frac{%
f(a)+f(b)}{2}  \label{eq:HermiteHadamard}
\end{equation}%
and given that the pdf of the standard normal $\phi (x)$ is convex for $x>1$
we obtain:%
\begin{equation*}
\Phi (b)-\Phi (a)=\int_{a}^{b}\phi (x)~dx\geq (b-a)\phi \left( \frac{a+b}{2}%
\right) ,\ \ \ \ b>a>1
\end{equation*}%
So for $z_{2}>\chi _{\alpha }^{2}>z_{1}>1$ we can write for the four terms
in (\ref{eq:LowerNumerator4terms}) 
\begin{eqnarray}
G(z_{2};\lambda )-G(z_{1};\lambda ) &\geq &(\sqrt{z_{2}}-\sqrt{z_{1}})\left[
\phi \left( \frac{2\sqrt{\lambda }-\sqrt{z_{1}}-\sqrt{z_{2}}}{2}\right)
+\phi \left( \frac{2\sqrt{\lambda }+\sqrt{z_{1}}+\sqrt{z_{2}}}{2}\right) %
\right]   \notag \\
&>&(\sqrt{z_{2}}-\sqrt{z_{1}})\phi \left( \frac{2\sqrt{\lambda }-\sqrt{z_{1}}%
-\sqrt{z_{2}}}{2}\right) ,  \label{eq:LowerNumerator}
\end{eqnarray}%
since $\phi (\sqrt{\lambda }+1/2\sqrt{z_{1}}+1/2\sqrt{z_{2}})>0$. When $%
\epsilon $ is small, $z_{1}$ and $z_{2}$ are very similar and the lower
bound based on (\ref{eq:HermiteHadamard}) will be tight. \newline
For the upper bound of the denominator $G(\chi _{\alpha }^{2};\lambda )$ we
use the following inequalities:%
\begin{equation}
L_{1}(x)=1-\phi (x)\frac{1}{x}\leq \Phi (x)\leq U(x)=1-\phi (x)\frac{x}{%
x^{2}+1},  \label{eq:Aggarwal19}
\end{equation}%
see, e.g. \citet[Eq. (15)]{Aggarwal2019} in Equation (\ref{eq:AlternativeG}):%
\begin{eqnarray}
G\left( \chi _{\alpha }^{2};\lambda \right)  &\leq &\Phi (\sqrt{\lambda }+%
\sqrt{\chi _{\alpha }^{2}})-L_{1}(\sqrt{\lambda }-\sqrt{\chi _{\alpha }^{2}})
\notag \\
&<&1-L_{1}(\sqrt{\lambda }-\sqrt{\chi _{\alpha }^{2}})=\phi (\sqrt{\lambda }-%
\sqrt{\chi _{\alpha }^{2}})\frac{1}{\sqrt{\lambda }-\sqrt{\chi _{\alpha }^{2}%
}}.  \label{eq:UpperDenominator}
\end{eqnarray}%
Combining the bounds (\ref{eq:LowerNumerator}) and (\ref{eq:UpperDenominator}%
) gives:%
\begin{eqnarray}
\frac{G(z_{2};\lambda )-G(z_{1};\lambda )}{G(\chi _{\alpha }^{2};\lambda )}
&>&(\sqrt{\lambda }-\sqrt{\chi _{\alpha }^{2}})(\sqrt{z_{2}}-\sqrt{z_{1}}%
)\times   \notag \\
&&\exp \left( -\tfrac{1}{8}(\sqrt{z_{1}}+\sqrt{z_{2}})^{2}+\tfrac{1}{2}\chi
_{\alpha }^{2}+\left( \tfrac{\sqrt{z_{1}}+\sqrt{z_{2}}}{2}-\sqrt{\chi
_{\alpha }^{2}}\right) \sqrt{\lambda }\right) .  \label{eq:InequalityRatio}
\end{eqnarray}%
The scaling factor in front of the exponential function is positive and
tends to infinity as $\lambda \rightarrow \infty $ since $z_{2}>\chi
_{\alpha }^{2}>z_{1}$. The limit behavior of the exponential function is
determined by the coefficient of $\sqrt{\lambda }$, which is positive (and
real) since%
\begin{equation*}
\frac{\sqrt{z_{1}}+\sqrt{z_{2}}}{2}>\sqrt{\chi _{\alpha }^{2}},
\end{equation*}%
where $\sqrt{z_{1}},\sqrt{z_{2}}$ and $\sqrt{\chi _{\alpha }^{2}}$ are the
quantiles of a standard normal distribution for $1-(\alpha +\epsilon )/2,$ $%
1-(\alpha -\epsilon )/2,$ $1-\alpha /2$ respectively and the quantile
function of a standard normal is convex for probabilities larger than 1/2.
This implies that the ratio in (\ref{eq:Ratio}) can be made arbitrarily
large by choosing $\lambda $ sufficiently large. This violates the condition
that it should be smaller than $\alpha /\epsilon .$ Hence there is a $%
\lambda _{0}$ such that $NRP(\lambda )>\alpha $ for all $\lambda
_{0}<\lambda <\infty $ and the test is over-sized for finite $\lambda $. Of
course, $\lim_{\lambda \rightarrow \infty }NRP(\lambda )=\alpha $ as shown
in Proposition 7 of the paper.

\subsection{Proof of Theorem 3}

We now consider the truncated-simply-augmented LR test defined by
augmentation of $CR_{LR}$ with the region $w_{b}^{3}$. This region is
defined as:%
\begin{eqnarray*}
w_{b}^{3} &=&\{(v_{1},v_{2}):bv_{2}<v_{1}<v_{2},0<v_{2}<\chi _{\alpha }^{2}\}
\\
&=&\{(v_{1},v_{2}):(0<v_{1}<b\chi _{\alpha }^{2},v_{1}<v_{2}<v_{1}/b)\cup
(b\chi _{\alpha }^{2}<v_{1}<\chi _{\alpha }^{2},v_{1}<v_{2}<\chi _{\alpha
}^{2})\}
\end{eqnarray*}%
for $0<b\leq 1$.

The probability content of region $w_{b}^{3}$ is given by:

\begin{eqnarray*}
\Pr [w_{b}^{3}] &=&\int_{0}^{\chi _{\alpha
}^{2}}\int_{bv_{2}}^{v_{2}}g(v_{1})g(v_{2};\lambda )+g(v_{2})g(v_{1};\lambda
)dv_{1}dv_{2} \\
&=&\int_{0}^{\chi _{\alpha
}^{2}}\int_{bv_{2}}^{v_{2}}g(v_{1})g(v_{2};\lambda )dv_{1}dv_{2} \\
&&+\int_{0}^{b\chi _{\alpha
}^{2}}\int_{v_{1}}^{v_{1}/b}g(v_{2})g(v_{1};\lambda
)dv_{2}dv_{1}+\int_{b\chi _{\alpha }^{2}}^{\chi _{\alpha
}^{2}}\int_{v_{1}}^{\chi _{\alpha }^{2}}g(v_{2})g(v_{1};\lambda )dv_{2}dv_{1}
\\
&=&\int_{0}^{\chi _{\alpha }^{2}}[G(v_{2})-G(bv_{2})]g(v_{2};\lambda )dv_{2}
\\
&&+\int_{0}^{b\chi _{\alpha }^{2}}[G(v_{1}/b)-G(v_{1})]g(v_{1};\lambda
)dv_{1}+\int_{b\chi _{\alpha }^{2}}^{\chi _{\alpha }^{2}}[G(\chi _{\alpha
}^{2})-G(v_{1})]g(v_{1};\lambda )dv_{1} \\
&=&\int_{0}^{b\chi _{\alpha }^{2}}[G(v)-G(bv)]g(v;\lambda )dv+\int_{b\chi
_{\alpha }^{2}}^{\chi _{\alpha }^{2}}[G(v)-G(bv)]g(v;\lambda )dv \\
&&+\int_{0}^{b\chi _{\alpha }^{2}}[G(v/b)-G(v)]g(v;\lambda )dv+\int_{b\chi
_{\alpha }^{2}}^{\chi _{\alpha }^{2}}[(1-\alpha )-G(v)]g(v;\lambda )dv \\
&=&\int_{0}^{b\chi _{\alpha }^{2}}[G(v/b)-G(bv)]g(v;\lambda )dv+\int_{b\chi
_{\alpha }^{2}}^{\chi _{\alpha }^{2}}[(1-\alpha )-G(bv)]g(v;\lambda )dv.
\end{eqnarray*}

Note the change in the order of integration in the third line. Since the
probability of $CR_{LR}$ is given by $\alpha -\alpha G(\chi _{\alpha
}^{2},\lambda )$, the difference between level $\alpha $ and the NRP is
given by $-\alpha G(\chi _{\alpha }^{2},\lambda )$. Noting that $G(\chi
_{\alpha }^{2},\lambda )=\int_{0}^{b\chi _{\alpha }^{2}}g(v;\lambda
)dv+\int_{b\chi _{\alpha }^{2}}^{\chi _{\alpha }^{2}}g(v;\lambda )dv$, the
difference between level $\alpha $ and the NRP\ of the truncated-augmented
critical region is given by the following discrepancy function:%
\begin{eqnarray}
\bar{D}_{\alpha }(b,\lambda ) &=&P[w_{b}]-\alpha G(\chi _{\alpha
}^{2},\lambda )  \notag \\
&=&\int_{0}^{b\chi _{\alpha }^{2}}[G(v/b)-G(bv)-\alpha ]g(v;\lambda
)dv+\int_{b\chi _{\alpha }^{2}}^{\chi _{\alpha }^{2}}[(1-\alpha
)-G(bv)-\alpha ]g(v;\lambda )dv  \notag \\
&=&\int_{0}^{b\chi _{\alpha }^{2}}f_{1}(v;b)g(v;\lambda )dv+\int_{b\chi
_{\alpha }^{2}}^{\chi _{\alpha }^{2}}f_{2}(v;b)g(v;\lambda )dv,
\label{eq:DescrepFunc}
\end{eqnarray}%
where the two functions $f_{1}(\cdot )$ and $f_{2}(\cdot )$ are independent
from $\lambda $. For a given $\lambda $, $P[w_{b}^{3}]$ increases as $b$
decreases. Hence, for each $\alpha $, there is an optimal value $b^{\ast }$
such that $D_{\alpha }(b^{\ast },\lambda )$ is numerically close to 0 for
some value of $\lambda $. Using the result%
\begin{equation}
\tfrac{\partial }{\partial \lambda }g(v;\lambda )=\tfrac{1}{2}%
[g_{3}(v;\lambda )-g(v;\lambda )]  \label{eq:Cohen88}
\end{equation}%
of \citet{Cohen1988}, the derivative of $\bar{D}_{\alpha }(b,\lambda )$ with
respect to $\lambda $ is given by:%
\begin{equation*}
\tfrac{\partial }{\partial \lambda }\bar{D}_{\alpha }(b,\lambda )=\tfrac{1}{2%
}\int_{0}^{b\chi _{\alpha }^{2}}f_{1}(v;b)[g_{3}(v;\lambda )-g(v;\lambda
)]dv+\tfrac{1}{2}\int_{b\chi _{\alpha }^{2}}^{\chi _{\alpha
}^{2}}f_{2}(v;b)[g_{3}(v;\lambda )-g(v;\lambda )]dv.
\end{equation*}%
For $\alpha =.05$, an iterative procedure between choosing $b$ and finding
the maximum value of $\bar{D}_{.05}(b,\lambda )$ gives $b^{\ast
}=0.86978984806$, where the maximum is obtained at $\lambda ^{\ast
}=3.844989224948$ such that $\bar{D}_{.05}(b^{\ast },\lambda ^{\ast
})=-2.95336772\cdot 10^{-13}$. The figure below shows $\bar{D}_{.05}(b^{\ast
},\lambda )$ for $1\leq \lambda \leq 25$ (left) and a close-up for $\lambda $
around $\lambda ^{\ast }$ (right).

\includegraphics[width=15cm]{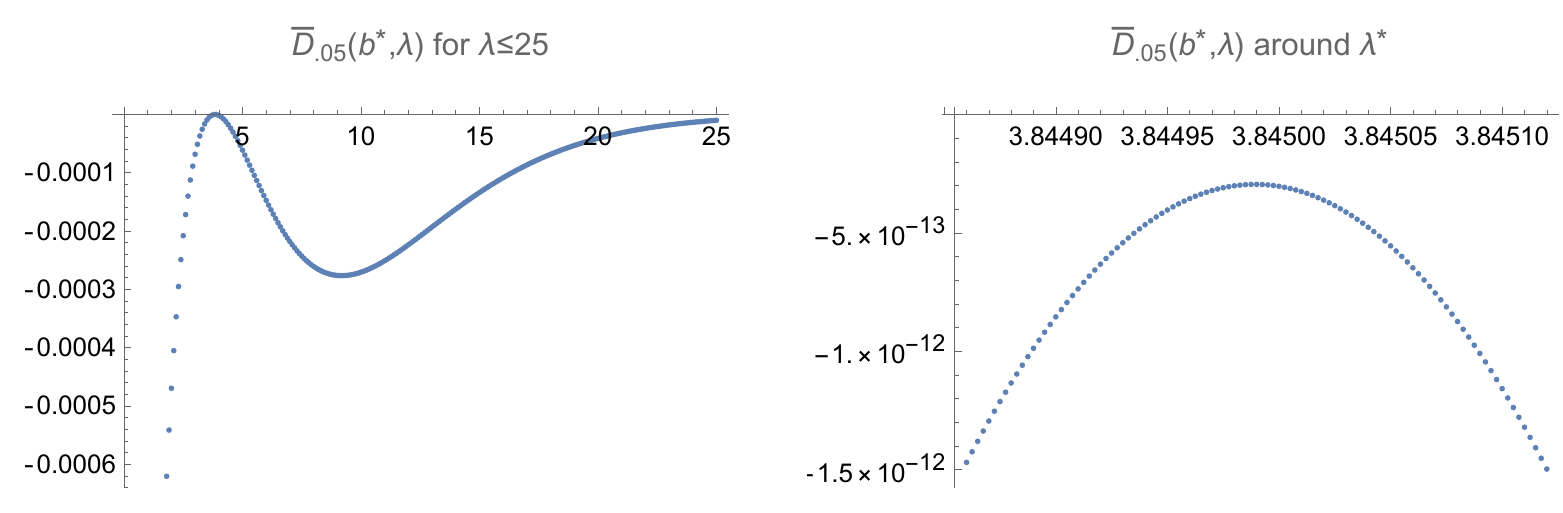}

If $\bar{D}_{.05}(b^{\ast },\lambda )\leq 0$ for all $\lambda \geq 0$, then
the test based on $CR_{LR}$ augmented with region $w_{b}^{3}$ is a valid $.05
$-level test. The figure above numerically shows that $\bar{D}_{.05}(b^{\ast
},\lambda )<0$ for $\lambda \leq 25$, but this does not guarantee that $\bar{%
D}_{.05}(b^{\ast },\lambda )$ remains below zero for larger values of $%
\lambda $. In Theorem 3 of the paper, it is claimed that $\bar{D}_{\alpha
}(b^{\ast },\lambda )\leq 0$, so $\bar{D}_{\alpha }(b^{\ast },\lambda )$ is
bounded from above by a function that becomes negative for sufficiently
large $\lambda (\alpha )$. This bounding function is based on an upper bound
of the two integrals in (\ref{eq:DescrepFunc}) in such a way that relatively
simple expressions in the pdf of the standard normal distribution $\phi
(\cdot )$ are obtained. In the figure below, $f_{1}(v;b^{\ast })$ and $%
f_{2}(v;b^{\ast })$ are shown in blue and orange respectively for $\alpha
=.05$.

\includegraphics[scale=1.0]{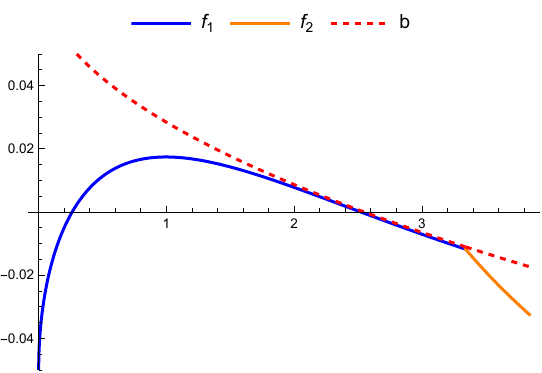}

We are now in the position to prove the following claim that implies Theorem
3 of the paper: \bigskip 

\emph{For }$\alpha <0.5$\emph{, there is a bounding function }$ub_{\alpha
}(\lambda )\leq 0$\emph{\ and a value }$\lambda _{0}(\alpha )$\emph{\ such
that }$D_{\alpha }(b^{\ast },\lambda )\leq ub_{\alpha }(\lambda )$\emph{\
for }$\lambda _{0}(\alpha )<\lambda <\infty $\emph{. \bigskip }

To show that the sum of the integrals in (\ref{eq:DescrepFunc}) is negative, 
$f_{1}(v;b^{\ast })$ and $f_{2}(v;b^{\ast })$ are bounded from above. Note
that for large values of $\lambda $, the pdf of the noncentral chi-square
behaves very similar to the pdf of a scaled standard normal, i.e.%
\begin{equation*}
g(v,\lambda )\sim \frac{1}{2\sqrt{v}}\phi (\sqrt{v}-\sqrt{\lambda }),
\end{equation*}%
which is an exponentially increasing function in (the scalar) $v$. As $%
\lambda $ increases, more weight is given to larger $v$-values. For $\alpha
=.05$, the bounding function is shown as the dashed red line in the figure
and corresponds to the function $\gamma _{0}+\gamma _{1}\sqrt{v}$ with $%
(\gamma _{0},\gamma _{1})=(0.07823,-0.04917)$. Note that the bound is only
tight for larger values of $v$, but this is sufficient as this area gets the
highest weight. The bounding function is chosen because the antiderivative
of $\sqrt{v}g(v;\lambda )$ is relatively simple:%
\begin{eqnarray}
\int \sqrt{v}g(v;\lambda )dv &=&-\sqrt{\lambda }G(\lambda ,z)-2vg(z,\lambda )
\notag \\
&=&-\sqrt{\lambda }[\Phi (\sqrt{\lambda }+\sqrt{v})+\Phi (\sqrt{\lambda }-%
\sqrt{v})-1]  \notag \\
&&-[\phi (\sqrt{v}-\sqrt{\lambda })+\phi (\sqrt{v}+\sqrt{\lambda })].
\label{eg:Integral1}
\end{eqnarray}%
Note that in the first line, the arguments of $G(\cdot )$ are switched,
which is done intentionally and is not a typo. Hence, the integrals in (\ref%
{eq:DescrepFunc}) are bounded by:%
\begin{equation}
\int_{0}^{\chi _{\alpha }^{2}}(\gamma _{0}+\gamma _{1}\sqrt{v})g(v;\lambda
)dv=\gamma _{0}G[z,\lambda ]+\gamma _{1}\int_{0}^{\chi _{\alpha }^{2}}\sqrt{v%
}g(v;\lambda )dv.  \label{eq:Integral2}
\end{equation}%
Substitution of (\ref{eg:Integral1}) in (\ref{eq:Integral2}) leads to%
\begin{eqnarray}
bound_{\alpha }(\lambda ) &=&\int_{0}^{\chi _{\alpha }^{2}}(\gamma
_{0}+\gamma _{1}\sqrt{v})g(v;\lambda )dv=\gamma _{0}[\Phi (\sqrt{\lambda }+%
\sqrt{\chi _{\alpha }^{2}})-\Phi (\sqrt{\lambda }-\sqrt{\chi _{\alpha }^{2}}%
)]  \label{eq:u1} \\
&&-\gamma _{1}\sqrt{\lambda }\left( \Phi (\sqrt{\lambda }+\sqrt{\chi
_{\alpha }^{2}})-\Phi (\sqrt{\lambda })-[\Phi (\sqrt{\lambda })-\Phi (\sqrt{%
\lambda }-\sqrt{\chi _{\alpha }^{2}})]\right)  \label{eq:u2} \\
&&-\gamma _{1}\left( \phi (\sqrt{\chi _{\alpha }^{2}}-\sqrt{\lambda })-\phi (%
\sqrt{\lambda })+\phi (\sqrt{\chi _{\alpha }^{2}}+\sqrt{\lambda })-\phi (%
\sqrt{\lambda })\right) ,  \label{eq:u3}
\end{eqnarray}%
where differences $\Phi (b)-\Phi (a)$ are such that $b>a$. In order to
investigate the behavior of $bound(\lambda )$ as $\lambda $ becomes large,
each of the three terms within parenthesis in (\ref{eq:u1})-(\ref{eq:u3})
will be bounded. For the bounding function, it is important to note that $%
\gamma _{0}>0$, whereas $\gamma _{1}<0$.

For the first two terms, the inequalities shown in (\ref{eq:Aggarwal19}) are
used. Using the upper bound for $\Phi (\sqrt{\lambda }+\sqrt{\chi_{\alpha
}^{2}})$ and the lower bound for $\Phi (\sqrt{\lambda }-\sqrt{\chi_{\alpha
}^{2}})$, the first term in (\ref{eq:u1}) is bounded by%
\begin{eqnarray*}
\Phi (\sqrt{\lambda }+\sqrt{\chi_{\alpha }^{2}})-\Phi (\sqrt{\lambda }-\sqrt{%
\chi_{\alpha }^{2}}) &\leq &U(\sqrt{\lambda }+\sqrt{\chi_{\alpha }^{2}}%
)-L_{1}(\sqrt{\lambda }-\sqrt{\chi_{\alpha }^{2}}) \\
&=&\frac{\phi (\sqrt{\lambda }-\sqrt{\chi_{\alpha }^{2}})}{\sqrt{\lambda }-%
\sqrt{\chi_{\alpha }^{2}}}-\frac{\phi (\sqrt{\lambda }+\sqrt{\chi_{\alpha
}^{2}})(\sqrt{\lambda }+\sqrt{\chi_{\alpha }^{2}})}{(\sqrt{\lambda }+\sqrt{%
\chi_{\alpha }^{2}})^{2}+1}=ub_{1,\alpha }(\lambda ).
\end{eqnarray*}%
Taking appropriate upper and lower bounds, the second term is bounded by%
\begin{eqnarray*}
\Phi (\sqrt{\lambda }+\sqrt{\chi_{\alpha }^{2}})-\Phi (\sqrt{\lambda }%
)-[\Phi (\sqrt{\lambda })-\Phi (\sqrt{\lambda }-\sqrt{\chi_{\alpha }^{2}})]
&\leq &U(\sqrt{\lambda }+\sqrt{\chi_{\alpha }^{2}})-L_{1}(\Phi (\sqrt{%
\lambda })) \\
&&-[L_{1}(\Phi (\sqrt{\lambda }))-U(\sqrt{\lambda }-\sqrt{\chi_{\alpha }^{2}}%
)]=ub_{2,\alpha }(\lambda ).
\end{eqnarray*}%
The last term is the smallest in magnitude since it does not involve
integrals over $\phi $. For $\lambda >\chi_{\alpha }^{2}/4$, we have $\phi (%
\sqrt{\chi_{\alpha }^{2}}-\sqrt{\lambda })>\phi (\sqrt{\lambda })>\phi (%
\sqrt{\chi_{\alpha }^{2}}+\sqrt{\lambda })$, so the first term in (\ref%
{eq:u3}) can be bounded by:%
\begin{equation*}
\phi (\sqrt{\chi_{\alpha }^{2}}-\sqrt{\lambda })-\phi (\sqrt{\lambda })+\phi
(\sqrt{\chi_{\alpha }^{2}}+\sqrt{\lambda })-\phi (\sqrt{\lambda })<\phi (%
\sqrt{\chi_{\alpha }^{2}}-\sqrt{\lambda })=ub_{3,\alpha }(\lambda ).
\end{equation*}

Using the three upper bounds, the total can be bounded from above by%
\begin{equation*}
bound_{\alpha }(\lambda )<\gamma _{0}ub_{1,\alpha }(\lambda )-\gamma
_{1}ub_{2,\alpha }(\lambda )-\gamma _{1}ub_{3,\alpha }(\lambda ,\alpha
)=ub_{\alpha }(\lambda ).
\end{equation*}%
Algebraic simplifications show that the upper bound $ub(\lambda )$ can be
written as%
\begin{equation}
ub_{\alpha }(\lambda )=\frac{\exp (-\lambda /2)}{\sqrt{2\pi }}\left( \exp (-%
\frac{\chi _{\alpha }^{2}}{2}+\sqrt{\chi _{\alpha }^{2}\lambda })c_{1,\alpha
}(\lambda \,)-\exp (-\frac{\chi _{\alpha }^{2}}{2}-\sqrt{\chi _{\alpha
}^{2}\lambda })c_{2,\alpha }(\lambda )-2\gamma _{1}\right) ,  \label{eq:up}
\end{equation}%
where%
\begin{eqnarray}
c_{1,\alpha }(\lambda ) &=&\frac{\gamma _{0}}{\sqrt{\lambda }-\sqrt{\chi
_{\alpha }^{2}}}+\gamma _{1}\frac{1-\chi _{\alpha }^{2}+\sqrt{\chi _{\alpha
}^{2}\lambda }}{1+\chi _{\alpha }^{2}-2\sqrt{\chi _{\alpha }^{2}\lambda }%
+\lambda }  \notag \\
&=&\frac{(1+\chi _{\alpha }^{2})(\gamma _{0}+\sqrt{\chi _{\alpha }^{2}}%
\gamma _{1})-(2\sqrt{\chi _{\alpha }^{2}}+\gamma _{1}+2z\gamma _{1})\sqrt{%
\lambda }+(\gamma _{0}+\sqrt{\chi _{\alpha }^{2}}\gamma _{1})\lambda }{-%
\sqrt{\chi _{\alpha }^{2}}(1+\chi _{\alpha }^{2})+(1+3\chi _{\alpha }^{2})%
\sqrt{\lambda }-3\sqrt{z}\lambda +\lambda ^{3/2}}  \label{eq:c1} \\
c_{2,\alpha }(\lambda ) &=&\frac{(\sqrt{\chi _{\alpha }^{2}}+\sqrt{\lambda }%
)(\gamma _{0}-\gamma _{1}\sqrt{\lambda })}{1+z+2\sqrt{\chi _{\alpha
}^{2}\lambda }+\lambda }.  \notag
\end{eqnarray}%
When $\gamma _{0}>0$ and $\gamma _{1}<0$, $c_{2,\alpha }(\lambda )>0$ for $%
\lambda >0$ and $\exp (-\chi _{\alpha }^{2}/2-\sqrt{\chi _{\alpha
}^{2}\lambda })>0$ tends exponentially to zero as $\lambda \rightarrow
\infty $. Hence, we see that the sign of $ub_{\alpha }(\lambda )$ is mainly
determined by the sum $\exp (-\chi _{\alpha }^{2}/2+\sqrt{\chi _{\alpha
}^{2}\lambda })c_{1,\alpha }(\lambda )-2\gamma _{1}$. Since $\exp (-\chi
_{\alpha }^{2}/2+\sqrt{\chi _{\alpha }^{2}\lambda })>0$ becomes
exponentially large as $\lambda $ increases, the sign of the sum will be
determined by the sign of $c_{1,\alpha }(\lambda )$. The coefficient $%
c_{1,\alpha }(\lambda )$ can be written as a ratio of two polynomials in $%
\sqrt{\lambda }$, which is done in (\ref{eq:c1}). Since%
\begin{equation*}
c_{1,\alpha }(\lambda )\sim \frac{(\gamma _{0}+\sqrt{\chi _{\alpha }^{2}}%
\gamma _{1})}{\sqrt{\lambda }},
\end{equation*}%
the function $c_{1,\alpha }(\lambda )$ will become and stay negative for
sufficiently large $\lambda $ when $\gamma _{0}<-\sqrt{\chi _{\alpha }^{2}}%
\gamma _{1}$. For selected values of $\alpha $, Table \ref%
{table:bstar_lambda0} shows the optimal values $b^{\ast }$, the value for
which the discrepancy function has a local maximum $\lambda ^{\ast }$ and
the numerical value of $(\gamma _{0},\gamma _{1})$ such that $(\gamma
_{0}+\gamma _{1}\sqrt{v})$ is larger than $f_{1}(v;b^{\ast })$ for $0\leq
v\leq b^{\ast }\chi _{\alpha }^{2}$ and $f_{2}(v;b^{\ast })$ for $b^{\ast
}\chi _{\alpha }^{2}\leq v\leq \chi _{\alpha }^{2}$. Furthermore, the last
column shows $\lambda _{0}(\alpha )$ such that $ub_{\alpha }(\lambda )$ is
below zero for $\lambda >\lambda _{0}(\alpha )$. For $\alpha =.05$, the
ratio in (\ref{eq:c1}) given by%
\begin{equation*}
c_{1,.05}(\lambda )=\frac{-0.087831+0.120283\sqrt{\lambda }-0.0181414\lambda 
}{-9.48908+12.5244\sqrt{\lambda }-5.87989\lambda +\lambda ^{3/2}},
\end{equation*}%
which turns negative for $\lambda >33.580,$ whereas Table \ref%
{table:bstar_lambda0} shows that $ub_{.05}(\lambda )<0$ for $\lambda >33.64$%
. Both functions converge to zero from below as $\lambda \rightarrow \infty
. $ The figure below shows $\bar{D}_{.05}(b^{\ast },\lambda )$ as well as
the upper bound $ub_{.05}(\lambda )$ in the region of $\lambda $ around $%
\lambda _{0}(.05)$.

\includegraphics[scale=1.0]{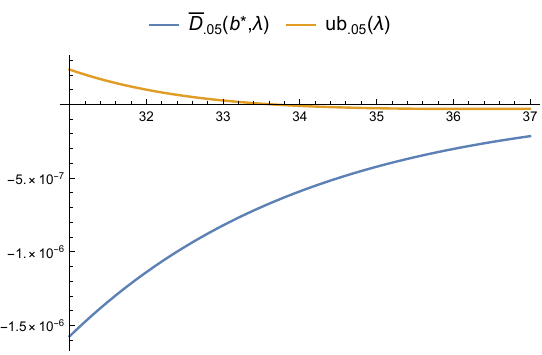}

\spacingset{1.0}%
\begin{table}[h]
\caption{Parameter values for the bounding function of the
truncated-simply-augmented LR test.}
\label{table:bstar_lambda0}
\begin{center}
\begin{tabular}{llllll}
$\alpha $ & $b^{\ast }$ & $\lambda ^{\ast }$ & $\gamma _{0}$ & $\gamma _{1}$
& $\lambda _{0}(\alpha )$ \\ \hline
0.01 & 0.969651 & 0.698449 & 0.01587 & -0.00917 & 22.252 \\ 
0.05 & 0.8697899 & 3.845 & 0.07823 & -0.04917 & 33.64 \\ 
0.1 & 0.7690223 & 7.7214 & 0.41014 & -0.28018 & 70.114 \\ 
0.2 & 0.5743807 & 13.6775 & 0.46939 & -0.41971 & 70.6 \\ 
0.3 & 0.3582853 & 18.5516 & 0.37195 & -0.42148 & 72.091 \\ 
0.4 & 0.1324455 & 24.483 & 0.19823 & -0.28368 & 82.639 \\ 
0.49 & 0.001961311 & 34.225 & 0.0200998 & -0.0353299 & 114.58 \\ \hline
\end{tabular}%
\end{center}
\end{table}
\clearpage
\spacingset{1.9}%

\subsection{Proof of Proposition 4}

Under $H_{0}$ the probability content of the augmenting region is given by%
\begin{eqnarray*}
\Pr [(v_{1},v_{2})\in w_{b};\lambda ] &=&\int_{0<v_{2}<\chi _{\alpha
}^{2}}\int_{bv_{2}<v_{1}<v_{2}}pdf(v_{1},v_{2};\lambda )dv_{1}dv_{2} \\
&&+\int_{\chi _{\alpha }^{2}<v_{2}<\chi _{\alpha
}^{2}/b}\int_{bv_{2}<v_{1}<\chi _{\alpha }^{2}}pdf(v_{1},v_{2};\lambda
)dv_{1}dv_{2}.
\end{eqnarray*}%
Substituting for the density and evaluating the integral over $v_{1}$
produces, after simplification,%
\begin{eqnarray*}
\Pr [(v_{1},v_{2})\in w_{b};\lambda ] &=&G(\chi _{\alpha }^{2})[G(\chi
_{\alpha }^{2}/b;\lambda )-G(\chi _{\alpha }^{2};\lambda )]+G(\chi _{\alpha
}^{2};\lambda )G(\chi _{\alpha }^{2}/b) \\
&&-\int_{0<v_{2}<\chi _{\alpha }^{2}/b}\left[ g(v_{2};\lambda
)G(bv_{2})+g(v_{2})G(bv_{2};\lambda )\right] dv_{2}.
\end{eqnarray*}%
Integrating each term in the second line by parts gives%
\begin{equation*}
\Pr [(v_{1},v_{2})\in w_{b};\lambda ]=b\int_{0<v_{2}<\chi _{\alpha }^{2}/b}
\left[ G(v_{2};\lambda )g(bv_{2})+G(v_{2})g(bv_{2};\lambda )\right]
dv_{2}-G(\chi _{\alpha }^{2})G(\chi _{\alpha }^{2};\lambda ).
\end{equation*}%
Then, transforming to $v=bv_{2}$ in the integral, we obtain 
\begin{equation*}
\Pr [(v_{1},v_{2})\in w_{b};\lambda ]=\int_{0<v<\chi _{\alpha }^{2}}\left[
G(v/b;\lambda )g(v)+G(v/b)g(v;\lambda )\right] dv-(1-\alpha )G(\chi _{\alpha
}^{2};\lambda ),
\end{equation*}%
and the result follows.

\subsection{Proof of Proposition 5}

Expanding the two noncentral components in the integrand in $A_{\alpha
}(b;\lambda )$ as Poisson mixtures we have 
\begin{equation*}
A_{\alpha }(b;\lambda )=e^{-\frac{1}{2}\lambda }\sum_{j=0}^{\infty }\frac{%
(\lambda /2)^{j}}{j!}\int_{0<v<\chi _{\alpha
}^{2}}[g_{2j+1}(v)G(v/b)+g(v)G_{2j+1}(v/b)]dv,
\end{equation*}%
and also 
\begin{equation*}
G(\chi _{\alpha }^{2};\lambda )=e^{-\frac{1}{2}\lambda }\sum_{j=0}^{\infty }%
\frac{(\lambda /2)^{j}}{j!}G_{2j+1}(\chi _{\alpha }^{2}).
\end{equation*}%
The two power series coincide for all $\lambda $ if\ and only if all
coefficients agree, that is 
\begin{equation*}
\int_{0<v<\chi _{\alpha
}^{2}}[g_{2j+1}(v)G(v/b)+g(v)G_{2j+1}(v/b)]dv=G_{2j+1}(\chi _{\alpha }^{2})
\end{equation*}%
for all $j.$ There is no $b\in (0,1]$ satisfying this equation for all $j.$

\subsection{Proof of Proposition 7}

It is well-known that $\lim_{\lambda \rightarrow \infty }G(z;\lambda )=0$
for any finite $z>0.$ The term $A_{\alpha }(b;\lambda )$ in $D_{\alpha
}(b,\lambda )$ is evidently positive, and is less than $G(\chi _{\alpha
}^{2}/b)G(\chi _{\alpha }^{2};\lambda )+G(\chi _{\alpha }^{2})G(\chi
_{\alpha }^{2}/b;\lambda )$ for all $\lambda .$ Since this converges to $0$
as $\lambda \rightarrow \infty $ for any $b>0,$ both terms in $D_{\alpha
}(b,\lambda )$ go to zero as $\lambda \rightarrow \infty .$

\clearpage\pagebreak

\section{Numerical determination of optimal b-value for the simply-augmented
LR test}

\begin{figure}[h]
\begin{center}
\includegraphics[width=1.1\textwidth]{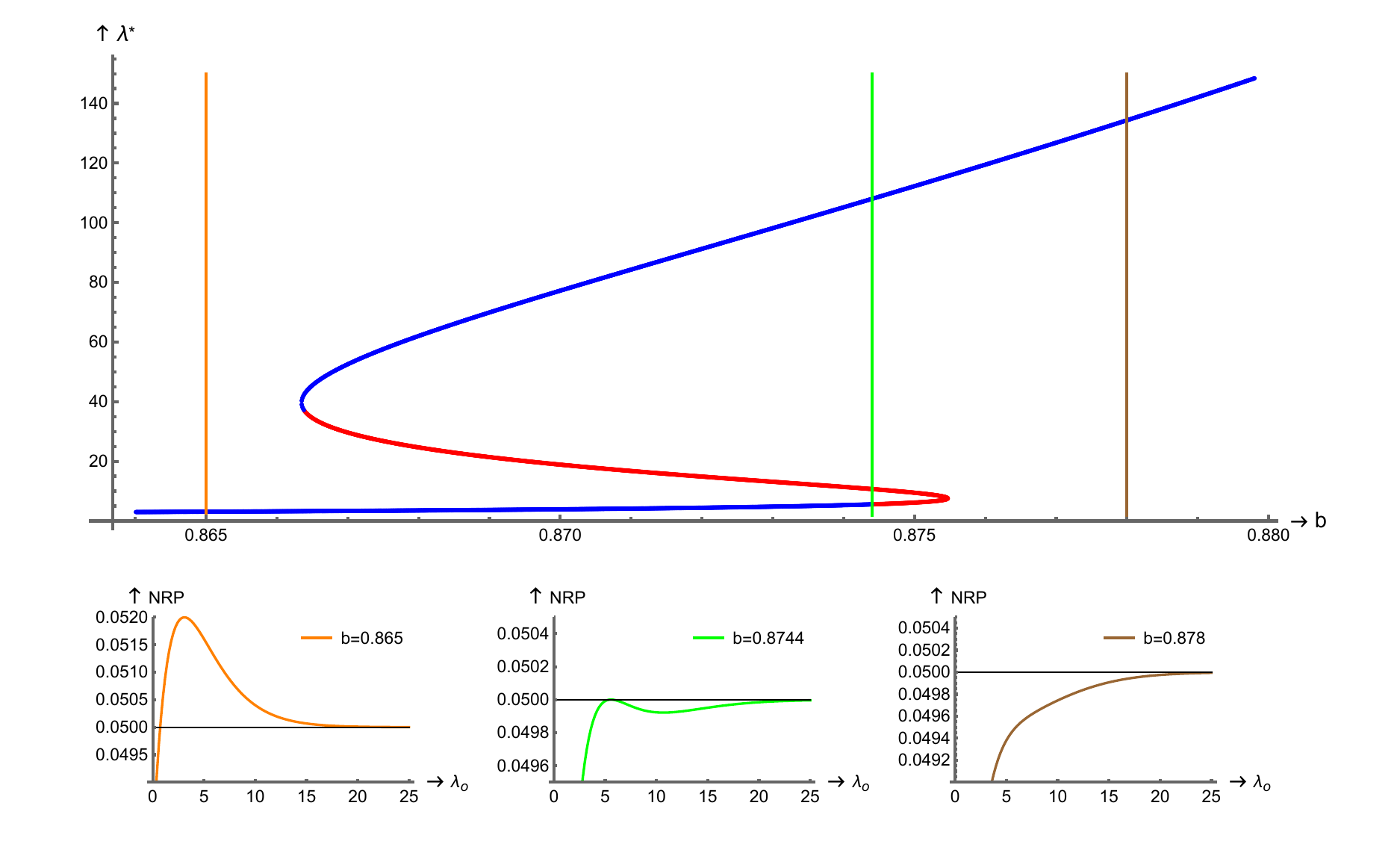}
\end{center}
\par
\vspace*{-6mm}
\caption{Top: the blue-red line traces the roots of the derivative of the
discrepancy $D_{.05}^{\prime }(b,\protect\lambda )$ for a given value of $b$%
. Blue/red indicates that the NRP for that given $(b,\protect\lambda ^{\ast
})$-value is smaller/larger than $\protect\alpha =.05$. Bottom: the
associated NRPs for three values of $b$, i.e. $b\in \{0.865,0.8744,0.878\}$.
These $b$-values are indicated by the three colored vertical lines in the
top figure. For $0.86635\leq b\leq 0.87547$, the NRP as function of $\protect%
\lambda $ has three stationary points (2 maxima and 1 minimum), while
outside this interval, the NRP only has one stationary point (1 maximum).}
\label{fig:Fig_b_vs_lambda_05}
\end{figure}

Below, an alternative expression is derived for the discrepancy function for
the simply-augmented $LR(b)$ test with the same structure as in (\ref%
{eq:DescrepFunc}). The probability content of region $w_{b}=w_{b}^{2}\cup
w_{b}^{3}$ is given by%
\begin{eqnarray*}
P_{w_{b}}(\lambda ) &=&\int_{0}^{\chi _{\alpha
}^{2}/b}\int_{bv_{2}}^{v_{2}}g(v_{1})g(v_{2};\lambda
)+g(v_{2})g(v_{1};\lambda )dv_{1}dv_{2} \\
&=&\int_{0}^{\chi _{\alpha
}^{2}}\int_{bv_{2}}^{v_{2}}g(v_{1})g(v_{2};\lambda )dv_{1}dv_{2}+\int_{\chi
_{\alpha }^{2}}^{\chi _{\alpha }^{2}/b}\int_{bv_{2}}^{\chi _{\alpha
}^{2}}g(v_{1})g(v_{2};\lambda )dv_{1}dv_{2} \\
&&+\int_{0}^{\chi _{\alpha
}^{2}}\int_{v_{1}}^{v_{1}/b}g(v_{2})g(v_{1};\lambda )dv_{2}dv_{1} \\
&=&\int_{0}^{\chi _{\alpha }^{2}}[G(v_{2})-G(bv_{2})]g(v_{2};\lambda
)dv_{2}+\int_{\chi _{\alpha }^{2}}^{\chi _{\alpha }^{2}/b}[(1-\alpha
)-G(bv_{2})]g(v_{2};\lambda )dv_{2} \\
&&+\int_{0}^{\chi _{\alpha }^{2}}[G(v_{1}/b)-G(v_{1})]g(v_{1};\lambda )dv_{1}
\\
&=&\int_{0}^{\chi _{\alpha }^{2}}[G(v/b)-G(bv)]g(v;\lambda )dv+\int_{\chi
_{\alpha }^{2}}^{\chi _{\alpha }^{2}/b}[(1-\alpha )-G(bv)]g(v;\lambda )dv.
\end{eqnarray*}%
The discrepancy function is given by $D_{\alpha }(b,\lambda
)=P_{w_{b}}(\lambda )+P_{A_{1}}(\lambda )-\alpha =P_{w_{b}}(\lambda )-\alpha
G(\chi _{\alpha }^{2};\lambda )$. Noting that $-\alpha G(\chi _{\alpha
}^{2};\lambda )=\int_{0}^{\chi _{\alpha }^{2}}-\alpha g(v;\lambda )dv$, we
obtain%
\begin{eqnarray*}
D_{\alpha }(b,\lambda ) &=&\int_{0}^{\chi _{\alpha
}^{2}}[G(v/b)-G(bv)-\alpha ]g(v;\lambda )dv+\int_{\chi _{\alpha }^{2}}^{\chi
_{\alpha }^{2}/b}[(1-\alpha )-G(bv)]g(v;\lambda )dv \\
&=&\int_{0}^{\chi _{\alpha }^{2}}h_{1}(v;b,\alpha )g(v;\lambda
)dv+\int_{\chi _{\alpha }^{2}}^{\chi _{\alpha }^{2}/b}h_{2}(v;b,\alpha
)g(v;\lambda )dv,
\end{eqnarray*}%
where the two functions $h_{1}(\cdot )$ and $h_{2}(\cdot )$ are independent
from $\lambda $. Using the result in (\ref{eq:Cohen88}), the derivative of $%
D_{\alpha }(b,\lambda )$ with respect to $\lambda $ is given by:%
\begin{equation}
D_{\alpha }^{\prime }(b,\lambda )=\tfrac{1}{2}\int_{0}^{\chi _{\alpha
}^{2}}h_{1}(v;b,\alpha )[g_{3}(v;\lambda )-g(v;\lambda )]dv+\tfrac{1}{2}%
\int_{\chi _{\alpha }^{2}}^{\chi _{\alpha }^{2}/b}f_{2}(v;b,\alpha
)[g_{3}(v;\lambda )-g(v;\lambda )]dv.  \label{dlambda_D_augmented}
\end{equation}%
When $b=1$, the area $w_{b}$ reduces to zero and the discrepancy function is
negative for any $\lambda \geq 0$, i.e. $D_{\alpha }(1,\lambda
)=P_{A_{1}}(\lambda )-\alpha =-\alpha G(\chi _{\alpha }^{2};\lambda )<0$.
For $b<1$, the discrepancy function can become positive. The goal is to
determine $b$ as small as possible given a maximum positive deviation. The
shape of the discrepancy function changes with $\alpha $ and \thinspace $b$.
For instance, for $\alpha =.05$, $D_{.05}(b,\lambda )$ has three stationary
points for $0.86635\leq b\leq 0.87547$, but only one stationary point
outside this interval given a grid of $b$-values. However, for $\alpha =.01$%
, $D_{.01}^{\prime }(b,\lambda )$ has one root and hence $D_{.01}(b,\lambda )
$ only possesses one maximum for $b<1$; see Figure \ref{fig:Fig_b_vs_lambda_05}
for $\alpha =.05$.

To find the smallest value of $b$ such that $D_{\alpha }(b,\lambda
)<\epsilon $ for $\epsilon \in \{10^{-9},10^{-16}\}$ the following algorithm
is used. For a given value of $b$, the number of sign changes in the
derivative of $D_{\alpha }(b,\lambda )$ is determined for a grid of $\lambda 
$-values, i.e. $\lambda \in \{0.0001:0.01:5\}\cup \{5.2:0.2:30\}\cup
\{31:1:150\}$ where ${\{a:b:c\}}$ is a regular grid between $a$ and $c$ with
grid spacing $b$. For each sign change, a root of $D_{\alpha }^{\prime
}(b,\lambda )$ is found by bisection using formula (\ref{dlambda_D_augmented}%
). The number of roots varies with $b$, and we identify two cases: case 1:
only one root $\lambda _{1}^{\ast }$ is found or case 2: three roots $%
\lambda _{1}^{\ast }\geq \lambda _{2}^{\ast }\geq \lambda _{3}^{\ast }$ are
found. In each step, $b$ is reduced by $\Delta =0.0001$, i.e. $b\leftarrow
b-\Delta $. This is stopped when, in case 1: $D_{\alpha }(b,\lambda
_{1}^{\ast })>\epsilon $ or, in case 2: $D_{\alpha }(b,\lambda _{1}^{\ast
})>\epsilon $ or $D_{\alpha }(b,\lambda _{3}^{\ast })>-\epsilon $. Next, $%
\Delta $ is divided by 10, i.e. $\Delta \rightarrow \Delta /10$ and it is
checked if for the slightly higher value of $b$, i.e. $b+\Delta $, in case
1: if $D_{\alpha }(b+\Delta ,\lambda _{1}^{\ast })<\epsilon $ or in case 2:
if $D_{\alpha }(b+\Delta ,\lambda _{1}^{\ast })<\epsilon $ and $D_{\alpha
}(b+\Delta ,\lambda _{3}^{\ast })<-\epsilon $. When this is not the case,
repeat the procedure with an appropriate starting value, i.e. $b+10\Delta $
and the reduced $\Delta $ obtained before. All computations were done in
Julia version 1.8.5, see \citet{Bezanson2017}, and verified in Mathematica
13.1, see \citet{Mathematica}. Table \ref{tab:alfa_b_lambda_D} shows the
results for $\alpha \in \{.01,.05,.1\}$, where for $\alpha =.01$ the largest
root is determined outside the mentioned grid of $\lambda $ values using
arbitrary precision arithmetic in Julia using \texttt{setprecision(96)}
corresponding to approximately 32 significant digits.

\spacingset{1.0}%
\begin{table}[h]
\caption{The values $b(\protect\alpha )$, the roots $\protect\lambda %
_{j}^{\ast }$ of $D_{\protect\alpha }(b(\protect\alpha ),\protect\lambda )$
and the values of $D_{\protect\alpha }(b(\protect\alpha ),\protect\lambda %
_{j}^{\ast })$. For $\protect\alpha =.1$ there is only one large root $%
\protect\lambda _{1}^{\ast }$ and the test is oversized by less than $%
10^{-16}.$}
\label{tab:alfa_b_lambda_D}\vspace*{-2mm}
\par
\begin{center}
{\small 
\begin{tabular}{llllllll}
$\alpha $ & $b(\alpha )$ & $\lambda _{1}^{\ast }$ & $\lambda _{2}^{\ast }$ & 
$\lambda _{3}^{\ast }$ & $D(b(\alpha ),\lambda _{1}^{\ast })$ & $D(b(\alpha
),\lambda _{2}^{\ast })$ & $D(b(\alpha ),\lambda _{3}^{\ast })$ \\ \hline
.01 & 0.9696632222091674 & \multicolumn{1}{r}{6978.64} & 7.12466 & 0.710853
& 6.4613e-1432 & -0.000996 & -8.7188e-17 \\ 
.05 & 0.874403978704909 & \multicolumn{1}{r}{108.029} & 10.674 & 5.562 & 
3.526e-21 & -7.845e-5 & -6.703e-17 \\ 
.1 & 0.829720 & \multicolumn{1}{r}{79.958} &  &  & 9.929e-17 &  &  \\ \hline
\end{tabular}%
}
\end{center}
\end{table}
\clearpage
\spacingset{1.9}%

\section{Power difference graphs}

\label{sec:appendix_B_additional_graph_table}

\begin{figure}[h]
\begin{center}
\includegraphics[width=0.90\textwidth]{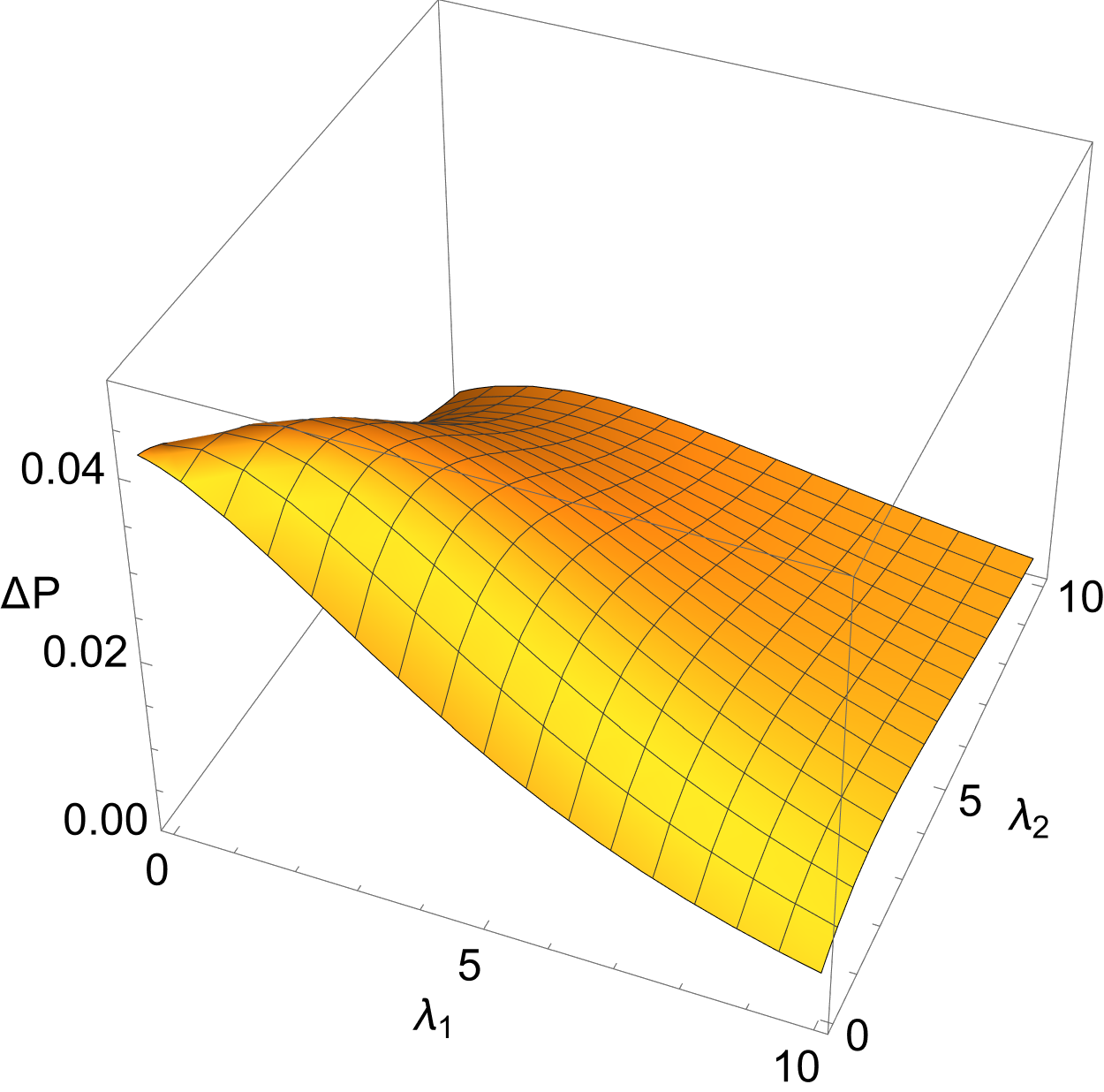}
\end{center}
\par
\vspace*{-6mm}
\caption{Power difference $(\Delta P)$ between the $LR(b)$ test and the $LR$
test; $\protect\alpha =.05.$ }
\label{fig:PowerDiff_LRvsAug}
\end{figure}

\begin{figure}[h]
\begin{center}
\includegraphics[width=0.90\textwidth]{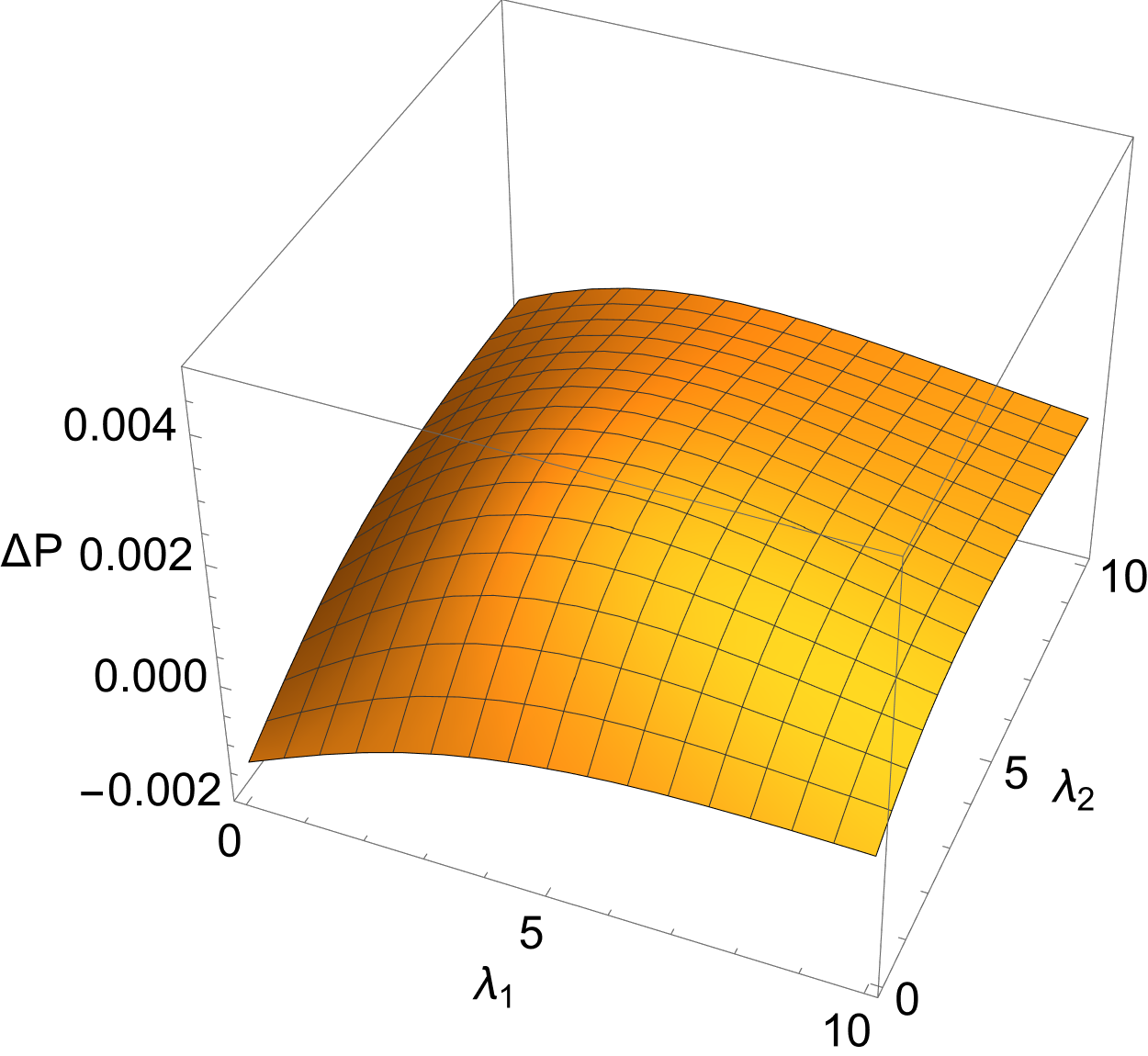}
\end{center}
\par
\vspace*{-6mm}
\caption{Power difference $(\Delta P)$ between the $LR(b)$ test and the $%
\overline{LR}(\overline{b})$ test; $\protect\alpha =.05.$ }
\label{fig:PowerDiff_TRvsAug}
\end{figure}

Note the difference in scale between Figure \ref{fig:PowerDiff_LRvsAug} and
Figure \ref{fig:PowerDiff_TRvsAug}.

\clearpage

\section{Additional simulation results: Heteroskedasticity}

\label{sec:appendix_D_additional_results}

\spacingset{1.0}%
\begin{table}[h]
\caption{Simulated NRPs ($\times 100\%$) for various variance
specifications: (i) $\protect\sigma (x_{i})=1$, (ii) $\protect\sigma %
(x_{i})=|x_{i}|$, and (iii) $\protect\sigma (x_{i})=\exp (0.4x_{i})$: $H_{0}:%
\protect\theta _{1}\protect\theta _{2}=0$ (with $\protect\theta _{1}=0$), $%
10^{6}$ replications. Left panel employs ordinary $t$-statistics, right
panel utilizes the robust $t$-statistics. }\vspace*{-3mm}
\par
\begin{center}
\begin{tabular}{lrrrrrrrrrrrrrrr}
& \multicolumn{7}{c}{Non-robust SE} &  & \multicolumn{7}{c}{Robust SE} \\ 
$n=50$ & \multicolumn{3}{c}{LR} &  & \multicolumn{3}{c}{$LR(b)$} &  & 
\multicolumn{3}{c}{LR} &  & \multicolumn{3}{c}{$LR(b)$} \\ 
$\theta _{2}$ & (i) & (ii) & (iii) &  & (i) & (ii) & (iii) &  & (i) & (ii) & 
(iii) &  & (i) & (ii) & (iii) \\ 
\cline{2-4}\cline{6-8}\cline{10-12}\cline{14-16}
0.0 & 0.3 & 3.8 & 1.6 &  & 4.6 & 7.3 & 5.5 &  & 0.5 & 0.8 & 0.7 &  & 4.6 & 
4.9 & 4.8 \\ 
0.14 & 0.9 & 5.6 & 2.8 &  & 4.8 & 8.8 & 6.4 &  & 1.2 & 1.3 & 1.4 &  & 5.0 & 
5.0 & 5.1 \\ 
0.39 & 4.0 & 13.8 & 8.3 &  & 5.2 & 15.4 & 9.8 &  & 4.8 & 4.2 & 4.9 &  & 6.0
& 6.4 & 6.8 \\ 
0.59 & 5.3 & 18.5 & 11.1 &  & 5.5 & 19.0 & 11.4 &  & 6.3 & 6.4 & 7.0 &  & 6.4
& 7.2 & 7.6 \\ \cline{2-4}\cline{6-8}\cline{10-12}\cline{14-16}
&  &  &  &  &  &  &  &  &  &  &  &  &  &  &  \\ 
$n=250$ & \multicolumn{3}{c}{LR} &  & \multicolumn{3}{c}{$LR(b)$} &  & 
\multicolumn{3}{c}{LR} &  & \multicolumn{3}{c}{$LR(b)$} \\ 
$\theta _{2}$ & (i) & (ii) & (iii) &  & (i) & (ii) & (iii) &  & (i) & (ii) & 
(iii) &  & (i) & (ii) & (iii) \\ 
\cline{2-4}\cline{6-8}\cline{10-12}\cline{14-16}
0.0 & 0.3 & 5.0 & 1.9 &  & 4.6 & 8.3 & 5.8 &  & 0.3 & 0.4 & 0.4 &  & 4.6 & 
4.6 & 4.6 \\ 
0.14 & 3.0 & 12.7 & 7.4 &  & 5.0 & 14.8 & 9.6 &  & 3.1 & 1.8 & 2.3 &  & 5.1
& 5.1 & 5.3 \\ 
0.39 & 5.1 & 22.4 & 13 &  & 5.1 & 22.4 & 13.0 &  & 5.3 & 5.5 & 5.7 &  & 5.3
& 5.6 & 5.7 \\ 
0.59 & 5.1 & 22.5 & 13 &  & 5.1 & 22.5 & 13.0 &  & 5.3 & 5.7 & 5.7 &  & 5.3
& 5.7 & 5.7 \\ \cline{2-4}\cline{6-8}\cline{10-12}\cline{14-16}
&  &  &  &  &  &  &  &  &  &  &  &  &  &  &  \\ 
$n=500$ & \multicolumn{3}{c}{LR} &  & \multicolumn{3}{c}{$LR(b)$} &  & 
\multicolumn{3}{c}{LR} &  & \multicolumn{3}{c}{$LR(b)$} \\ 
$\theta _{2}$ & (i) & (ii) & (iii) &  & (i) & (ii) & (iii) &  & (i) & (ii) & 
(iii) &  & (i) & (ii) & (iii) \\ 
\cline{2-4}\cline{6-8}\cline{10-12}\cline{14-16}
0.0 & 0.3 & 6.4 & 1.6 &  & 4.6 & 9.5 & 5.6 &  & 0.3 & 0.3 & 0.3 &  & 4.6 & 
4.6 & 4.7 \\ 
0.14 & 4.4 & 19.1 & 8.9 &  & 5.0 & 20.3 & 9.9 &  & 4.5 & 2.6 & 3.6 &  & 5.1
& 5.2 & 5.2 \\ 
0.39 & 5.1 & 25.3 & 11.0 &  & 5.1 & 25.3 & 11.0 &  & 5.2 & 5.4 & 5.3 &  & 5.2
& 5.4 & 5.3 \\ 
0.59 & 5.1 & 25.3 & 11.0 &  & 5.1 & 25.3 & 11.0 &  & 5.2 & 5.4 & 5.3 &  & 5.2
& 5.4 & 5.3 \\ \cline{2-4}\cline{6-8}\cline{10-12}\cline{14-16}
\end{tabular}%
\end{center}
\label{table:NRP_hetero50250500}
\end{table}
\spacingset{1.0}%

\clearpage

\bibliographystyle{chicago}
\bibliography{references}

\end{document}